\documentclass{emulateapj}
\usepackage{natbib,apjfonts}
\newcommand{\lya}{Ly$\alpha$}
\newcommand{\Zsol}{Z_{\odot}}
\newcommand{\zsol}{\,$Z_{\odot}$}      
\newcommand{\Lsol}{L_{\odot}}
\newcommand{\mdot}{\,$\dot{M}$}        
\newcommand{\Rsol}{R_{\odot}}
\newcommand{\Msol}{M_{\odot}}
\newcommand{\msol}{\,$M_{\odot}$}      
\newcommand{\kms}{\,km\,s$^{-1}$}      
\newcommand{\sol}{\odot}
\newcommand{\hii}{H{\scriptsize ~II}}

\journalinfo{Accepted for publication in the Astrophysical Journal}
\slugcomment{To appear in the Astrophysical Journal in November 2004}

\shorttitle{Spectral Modelling of Star-forming Regions in the UV}
\shortauthors{Rix et al.}

\begin{document}

\title{SPECTRAL MODELLING OF STAR-FORMING REGIONS IN THE ULTRAVIOLET:
  STELLAR METALLICITY DIAGNOSTICS FOR HIGH REDSHIFT
  GALAXIES\altaffilmark{1}}

\author{Samantha A. Rix\altaffilmark{2} and Max Pettini}
\affil{Institute of Astronomy, Madingley Road, Cambridge, CB3 0HA, UK}

\author{Claus Leitherer} 
\affil{Space Telescope Science Institute, 3700 San Martin Drive,
  Baltimore, MD 21218}

\author{Fabio Bresolin and Rolf-Peter Kudritzki}
\affil{Institute for Astronomy, 2680 Woodlawn Drive, Honolulu HI 96822}

\and

\author{Charles C. Steidel}

\affil{Palomar Observatory, Caltech 105--24, Pasadena, CA 91125}
\altaffiltext{1}{Based, in part, on data obtained at the 
W.M. Keck Observatory, which is operated as a scientific partnership 
among the California Institute of Technology, the
University of California, and NASA, and was made possible by the 
generous financial support of the W.M. Keck Foundation.}
\altaffiltext{2}{Current address: Isaac Newton Group, Apartado de
  Correos 321, 38700 Santa Cruz de La Palma, Spain.}

\begin{abstract}
  
  The chemical composition of high redshift galaxies is an important
  property which gives clues to their past history and future
  evolution and yet is difficult to measure with current techniques.
  In this paper we investigate new metallicity indicators, based upon
  the strengths of stellar photospheric features at rest-frame
  ultraviolet wavelengths.  By combining the evolutionary spectral
  synthesis code {\em Starburst99\/} with the output from the non-LTE
  model atmosphere code {\em WM-basic\/}, we have developed a code
  that can model the integrated ultraviolet stellar spectra of
  star-forming regions at metallicities between 1/20 and twice solar.
  We use our models to explore a number of spectral regions that are
  sensitive to metallicity and clean of other spectral features.  The
  most promising metallicity indicator is an absorption feature
  between 1935\,\AA\ and 2020\,\AA, which arises from the blending of
  numerous \ion{Fe}{3} transitions.  We compare our model spectra to
  observations of two well studied high redshift star-forming
  galaxies, MS1512--cB58 (a Lyman break galaxy at $z_{\rm em} =
  2.7276$), and Q1307--BM1163 (a UV-bright galaxy at $z_{\rm em} =
  1.411$).  The profiles of the photospheric absorption features
  observed in these galaxies are well reproduced by the models. In
  addition, the metallicities inferred from their equivalent widths
  are in good agreement with previous determinations based on
  interstellar absorption and nebular emission lines. Our new
  technique appears to be a promising alternative, or complement, to
  established methods which have only a limited applicability at high
  redshifts.

\end{abstract}


\section{INTRODUCTION}

It is now possible, using a variety of techniques, to assemble
well-defined and large samples of galaxies at redshifts spanning most
of the Hubble time, from the present to $z = 5$.  While many of the
global properties of these galaxies---such as their luminosity
function, clustering, and contribution to the star formation rate
density---have been, or are, in the process of being characterised,
there have been relatively few studies to date of their individual
properties, such as their stellar populations and chemical enrichment,
particularly at $z > 1$. The simple reason is that, at these
redshifts, the faintness of all but the brightest galaxies makes
spectroscopic observations challenging, even when using the largest
aperture telescopes in the world in conjunction with highly efficient
spectrographs. Nevertheless, a small number of detailed observations,
which have targeted the brightest members of the population, have
begun to provide some insight into the chemical properties of high
redshift galaxies.

A number of authors, including \cite{teplitz00}, \cite{kobulnicky00}
and \cite{pettini01}, have used nebular emission lines of [\ion
{O}{2}], [\ion {O}{3}], and H$\beta$ \citep[the $R_{23}$ method
of][]{pagel79} to measure the oxygen abundance of the ionized gas in
galaxies at $z \simeq 3$ selected with the Lyman break technique
\citep{steidel96}.  These studies have shown that the brightest of
these Lyman break galaxies (LBGs) appear to be neither supersolar nor
as metal poor as the damped \lya\ systems at comparable redshifts.
Unfortunately, a degeneracy in the $R_{23}$ calibration can lead to
uncertainties of up to one order of magnitude in the derived oxygen
abundance.  More recently \cite{shapley04} have extended this work to
using [\ion {N}{2}] and H$\alpha$ \citep[the $N2$ index discussed
by][]{pp04}, to show that galaxies at $z \simeq 2$ that are bright at
rest-frame optical wavelengths are also metal-rich, with near-solar
values of (O/H).  However, one drawback of methods based on nebular
emission lines from \hii\ regions is that, at these redshifts, most of
the lines used as abundance diagnostics fall at near-infrared
wavelengths. In this spectral region observations from the ground are
made difficult by a multitude of strong night sky emission lines and
by the limitations of current detector technology.

It is simpler, observationally, to perform optical rather than
infrared spectroscopy and thereby study the rest-frame ultraviolet
(UV) spectra of high redshift galaxies.  Such spectra consist of the
integrated light from the hot and luminous O and B stars in the galaxy
({\em the stellar spectrum\/}) on which are superimposed the resonant
absorption lines produced by the interstellar gas ({\em the
  interstellar spectrum\/}). By exploiting the large wavelength
coverage and high efficiency of the Echellette Spectrograph and Imager
(ESI) on the Keck telescope, \cite{pettini02} obtained a high
resolution and high signal-to-noise ratio (S/N) optical spectrum of
the gravitationally lensed LBG MS1512--cB58 at $z = 2.7276$.  From a
detailed analysis of the interstellar absorption lines, these authors
were able to determine, for the first time, the abundance pattern of
many elements in a Lyman break galaxy.  As well as indicating a
relatively high metallicity, $Z \sim 2/5 \Zsol$, their results suggest
a rapid timescale for the metal enrichment, of the order of a few
hundred million years.  Such in-depth studies are, however,
observationally very demanding and with present means can only be
carried out on a few exceptionally bright galaxies.

To overcome these difficulties we consider in this paper a novel
technique for abundance determinations based on the rest-frame UV {\em
  stellar\/} spectrum.  Such a technique has of course been applied
extensively to the analysis of the optical spectra of {\em old\/}
stellar populations, such as in elliptical galaxies, for the last
twenty years through the well-established system of `Lick indices'
\citep{burstein84,faber85}.  However, its possible extension to the UV
spectra of star-forming galaxies has not been addressed up to now.
And yet, even at relatively low resolution and S/N, the rest-frame UV
spectra of high redshift galaxies show a rich variety of stellar
features which are mostly photospheric blends of absorption lines, as
well as lines produced in stellar winds.  It is thus worthwhile
exploring to what extent the stellar spectrum can yield new
metallicity probes \citep[see, for example, the very recent work
by][]{keel04}.

As the integrated stellar spectrum from a whole galaxy has
contributions from many different types of stars, it is not
straightforward to extract information about the underlying stellar
population. Nevertheless, progress in this direction can be made by
employing spectral synthesis codes such as {\em Starburst99\/}
\citep[][ and references therein]{leitherer99}. Such codes consider a
particular star formation law and stellar mass function and use
theoretical evolutionary tracks to evolve the stellar population with
time. Employing empirical libraries of stellar spectra to sum the
contributions from individual stars, the codes can then predict the
evolution of the population's integrated UV spectrum. By comparing an
observed spectrum with those synthesized with a range of different
model parameters, it is then possible to constrain the properties of
the underlying young stellar population [see \cite{leitherer03_hst10}
for a review].

The original empirical libraries adopted by {\em Starburst99\/} were
compiled from observations of Galactic stars obtained by the {\em
  International Ultraviolet Explorer (IUE)} satellite.  In order to
assess the sensitivity of the stellar spectrum to metallicity, a lower
metallicity library was later appended following a {\em Hubble Space
  Telescope (HST)} programme to observe hot stars in the metal-poor
Large and Small Magellanic Clouds \citep{leitherer01}. The success of
this work demonstrated the potential of using spectral synthesis
modelling to determine the metallicity of stellar populations, as
discussed in more detail below (\S\ref{subsec:sensitivity_Z}).
However, Galactic and Magellanic Clouds stars only sample a narrow
range of metallicities.  A major drawback for extending the empirical
libraries of stellar spectra to other metallicities is that only
relatively nearby stars can be observed with {\em HST} at the required
spectral resolution and S/N. Even enlarging the libraries to include
other Local Group galaxies would be challenging and observationally
expensive while providing little increase in metallicity parameter
space. With no prospects in the foreseeable future for extending the
empirical libraries to higher and lower metallicities, it may appear
as if we have reached a limit to this type of analysis.

Here we propose that a solution to this problem is to replace the
empirical libraries in {\em Starburst99\/} with theoretical ones.  A
similar approach has been recently developed for the Lick indices by
\cite{thomas03a} and \cite{thomas03b}.  Besides increasing the range
of metallicities that can be investigated, theoretical spectra offer
the additional benefits of a wider wavelength coverage and the
possibility of altering the relative abundances of different elements,
as well as the overall degree of metal enrichment, thus catering for a
variety of star formation histories.  This flexibility may be
particularly important when dealing with galaxies in the distant past,
when the dominant mode of star formation may have been quite different
from that commonly encountered in the nearby universe.  Fortunately,
the increasing sophistication of modern hot star model atmosphere
codes makes it possible to predict the emergent UV spectra from stars
of different structural parameters and stellar wind properties.

We have therefore used the non-LTE line-blanketed model atmosphere
code {\em WM-basic} \citep{pauldrach01}, which includes the effects of
stellar winds and spherical atmospheric extension, to create
theoretical libraries of stellar spectra over a range of metallicities
(\S\ref{subsec:wmbasic}) and integrated these libraries into {\em
  Starburst99\/} (\S\ref{subsec:starburst99}).  The outputs of {\em
  Starburst99\/} models thus produced compare favorably with more
conventional models based on empirical libraries of stellar spectra
(\S\ref{subsec:comp_emp_spec}).  The coupling of state-of-the-art
codes from the fields of hot stars and starburst galaxies has allowed
us, for the first time, to investigate the sensitivity of the
integrated spectrum of a star-forming region to metallicity from $Z =
0.05 \Zsol$ to $2 \Zsol$ (\S\ref{sec:synthetic_spectra}). We consider
in particular three blends of photospheric lines from OB
stars---centered at 1370\,\AA, 1425\,\AA, and 1978\,\AA---and conclude
that the last one of these is probably the most suitable for stellar
abundance determinations.  In \S\ref{sec:lbg_comparison} we compare
our models with Keck observations of two bright star-forming galaxies
and find that the metallicities deduced from the stellar line indices
agree to within a factor of $\sim 2$ with the values measured in the
interstellar media of these galaxies using established techniques.  We
discuss the results and draw our conclusions in
\S\ref{sec:conclusions}.

\section{SPECTRAL MODELLING OF STAR-FORMING GALAXIES}

In this section we describe how we first generated theoretical
spectral libraries with {\em WM-basic\/} and then incorporated them
into {\em Starburst99\/}, in order to create a code capable of
modelling the UV spectral features of star-forming regions over a
range in metallicities. Note that although {\em WM-basic\/} permits a
free choice of metallicities, the evolutionary tracks and model
atmospheres employed by {\em Starburst99\/} are restricted to
metallicities of 2, 1, 0.4, 0.2 and 0.05\zsol. For this reason, we
created theoretical libraries at these five metallicities.

\subsection{Generation of the {\it WM-basic\/} Spectral Libraries}\label{subsec:wmbasic}

\subsubsection{The Non-LTE Line-blanketed Stellar Wind Code {\it WM-basic\/}}

In order to predict the emergent stellar flux at different
wavelengths, it is necessary to produce a realistic model of the
atmospheric structure of a star.  For hot, massive stars this requires
a substantial effort because the physical conditions in their
atmospheres are complex and very different from those considered in
standard models.  They are dominated by the influence of the radiation
field, which has an energy density larger than, or of the same order
as, the energy density of atmospheric matter.  This has two important
consequences.  First, severe departures from local thermodynamic
equilibrium (LTE) are induced, because radiative transitions between
ionic energy levels become much more important than those caused by
inelastic collisions between ions and electrons. As a result, we
encounter non-LTE effects in the entire atmosphere of massive hot
stars.  Second, hydrodynamic outflows of atmospheric matter (stellar
winds) are initiated by line absorption of photons transferring
outwardly directed momentum to the atmospheric plasma.  The velocity
fields of these outflows are mostly subsonic in the stellar
photospheres, where the weak photospheric lines are formed, but become
highly supersonic above the photosphere, where the stronger spectral
lines are formed. Both non-LTE and stellar winds affect the emergent
energy distributions, ionizing fluxes and line spectra significantly
and their effects need to be included in the spectral synthesis of hot
stars \citep[for a detailed discussion see, for
example,][]{kudritzki98}.

Developing a code that solves the full complex set of coupled
equations describing the physics of hot star atmospheres is a
challenging numerical problem.  These equations include the radiation
hydrodynamics of the stellar wind (with the radiative line force from
millions of spectral lines), the statistical rate equations (for
determining the occupation numbers of different ionic species), the
spherical radiative transfer equations for the spectral lines and the
continuum, the energy equation and additional processes such as X-rays
generated by shocks in the stellar wind outflow.  It is not surprising
that the development of such codes has become feasible only in recent
years with increased computing power, improvements in numerical
techniques, and expanded databases of atomic data. Examples of these
sophisticated new codes are {\em CMFGEN\/} \citep{hillier98}, {\em
  FASTWIND\/} \citep{santolaya-rey97,repolust04}, and {\em WM-basic\/}
\citep{pauldrach94,pauldrach01}.

For the present work we chose to use the {\em WM-basic\/} code which
has the important advantage of being extremely fast thanks to its use
of a very advanced escape probability algorithm for the line transfer
in expanding media. While this is an approximate treatment of the line
transfer problem, it works very well when modelling the UV spectra of
hot stars \citep{pauldrach01}.  For the population synthesis work
described here the speed of computation is crucial, since one needs to
synthesize and add together the many different stellar spectral types
that contribute to the integrated spectrum of a young stellar
population.

Before explaining in more detail how we used the code for our
purposes, we note here that {\em WM-basic\/} treats the wind as
homogeneous, stationary and spherically symmetric. Although winds are
known to be time-dependent and unstable, these assumptions seem to be
reasonable approximations when modelling time- and population-averaged
spectral properties \citep[see, e.g.,][]{puls93, kudritzki00}.  {\em
  WM-basic\/} also allows for the inclusion of the effects of the
ionizing emission from randomly distributed shocks in the stellar wind
outflow using the approach developed by \cite{feldmeier97}.  Taking
into account emission from shocks may be important for modelling
correctly strong wind lines from high ionization stages such as
\ion{O}{6}, \ion{O}{5}, \ion{N}{5}, and \ion{C}{4} \citep{pauldrach94,
  pauldrach01, taresch97}, but photospheric and low ionization wind
lines are unaffected by this process.  Since in this first step of our
work we will focus on the spectral synthesis of photospheric line
features, we decided not to include the effects of shock emission at
this stage.

%%%%%%%%%%%%%%%%%%%%%%%%%%%%%%%%%%%%%%%%%%%%%%%%%%%%%%%%%%%%%%%%%%%%%%
\begin{figure}
\includegraphics[width=\columnwidth]{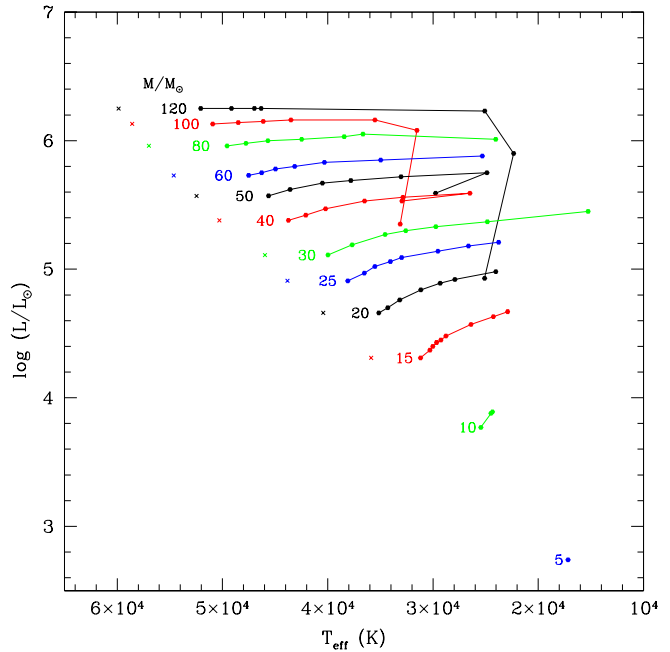}
\caption{Grid of {\em WM-basic\/} models. Models were computed {\em
for all five metallicities\/} at each point marked in the
Hertzsprung-Russell diagram. The grid points ({\em filled circles\/}) 
were chosen to lie on the solar metallicity
evolutionary tracks of \cite{meynet94}. The labels indicate the zero
age masses considered while the lines show the tracks 
along which each star evolves. 
On the main sequence, stars of lower metallicity than solar 
are hotter; we therefore considered additional 
points at higher temperatures ({\em crosses\/}) to sample this part of
the Hertzsprung-Russell diagram. The other stellar parameters used
when running the {\em WM-basic\/} models are summarised in
Table~\ref{tab:wmb_grid}.\label{fig:wmb_grid}}
\end{figure}
%%%%%%%%%%%%%%%%%%%%%%%%%%%%%%%%%%%%%%%%%%%%%%%%%%%%%%%%%%%%%%%%%%%%%%

\subsubsection{Computation of the Grid of Models}

{\em WM-basic\/} is distributed by A.W.A. Pauldrach as freeware
software\footnote{http://www.usm.uni-muenchen.de/people/adi/Programs/Download.html}
on the world wide web and can be run on an IBM-compatible personal
computer (PC) with a Microsoft Windows operating system, at least
350\,MHz CPU, 64\,MB RAM and 200\,MB free disk space. For the purposes
of this work, it was installed on a 2\,GHz Pentium PC in Cambridge and
on a 1\,GHz Athlon PC in Hawaii, both with 512\,MB RAM.  Despite the
sophistication of the code, the operation of {\em WM-basic\/} is
straightforward: the user specifies the stellar parameters of the star
to be modelled and the program outputs the corresponding synthetic
spectrum.  We computed theoretical UV spectra for 87 grid points in
the Hertzsprung-Russell (H-R) diagram at each of the five
metallicities selected, with an individual model taking between 35 and
80 minutes to run. As we required a high spectral resolution, for
improved accuracy, we requested output resolutions of 0.6\,\AA\ and
0.25\,\AA\ in the wavelength ranges 900--1150\,\AA\ and
1150--2124\,\AA\ respectively. In the remainder of this section we
discuss the selection of the grid points, which are summarised in
Table~\ref{tab:wmb_grid} and Figure~\ref{fig:wmb_grid}, and the other
input parameters used by {\em WM-basic\/}.

As the purpose of our grid of models was to create libraries of
theoretical spectra from which {\em Starburst99\/} could then
synthesise the integrated UV spectra of star-forming regions, we first
assessed which stars make a significant contribution to the total UV
light.  This is not immediately obvious because while hotter and more
luminous stars produce prodigious amounts of UV flux they are much
rarer and shorter-lived than lower mass stars. Simulations using {\em
  Starburst99\/} with its empirical libraries show that $\sim 90$\% of
the UV light at 1450\,\AA\ and 1800\,\AA\ is generated by stars with
zero age main sequence mass greater than 5\msol. For this reason, we
restricted our grid to stars evolving from ones with masses of at
least this value.

To run realistic spectral synthesis models one must sample this upper
part of the H-R diagram with sufficient grid points. As {\em
  Starburst99\/} employs the evolutionary tracks of \cite{meynet94} to
evolve the stellar population with time, we used these tracks as a
guide for which grid points to select. We chose points corresponding
to both zero age and evolved stars with initial masses of 120, 100,
80, 60, 50, 40, 30, 25, 20, 15, 10 and 5 \msol, as illustrated in
Figure~\ref{fig:wmb_grid}.  The evolutionary tracks were used to
specify the full set of stellar parameters at each grid point, namely
luminosity $L$, effective temperature $T_{\rm eff}$ and photospheric
radius $R$. We list these values in Table~\ref{tab:wmb_grid}.

It is also necessary to specify the stellar wind properties.  For
simplicity we first discuss stars of solar metallicity.  The terminal
velocity $v_{\infty}$ of the wind of each star was determined from the
empirical scaling formula of \cite{kudritzki00} which relates
$v_{\infty}$ to the effective temperature and photospheric escape
velocity as follows:
\begin{equation}
v_{\infty}=C(T_{\rm eff})\,v_{\rm esc},\quad {\rm where}
\end{equation}
\begin{eqnarray}
&C(T_{\rm eff})=2.65,\quad & T_{\rm eff}\ge 21,000\,{\rm K} \nonumber \\
&C(T_{\rm eff})=1.4,\quad & 10,000{\rm K}<T_{\rm eff}<21,000\,{\rm K} \nonumber\\
&C(T_{\rm eff})=1.0,\quad & T_{\rm eff}\le 10,000\,{\rm K}.\nonumber
\end{eqnarray}
Note that the photospheric escape velocity is defined as
\begin{equation}
v_{\rm esc}=[2gR (1-\Gamma) ]^{0.5}
\end{equation}
where $g$ is the photospheric gravity, $R$ is the photospheric radius
and $\Gamma$ is the ratio of radiative Thomson to gravitational
acceleration. This last parameter takes into account the effect of
Thomson scattering in weakening the effect of the gravitational
potential.

We also had to estimate the mass loss rate $\dot{M}$ from each star in
the grid. One of the predictions of radiation-driven wind theory [see
\cite{kudritzki98} for details] is that a wind momentum--luminosity
relation exists such that
\begin{equation}
\log D_{\rm mom}=\log D_{\rm 0}+x\log(L/\Lsol),
\end{equation}
for
\begin{displaymath}
D_{\rm mom}=\dot{M}v_{\infty}\,(R_{*}/\Rsol)^{0.5},
\end{displaymath}
where the coefficients $D_0$ and $x$ vary with spectral type and (for
O stars) with luminosity class.  By using the empirical calibrations
by \cite{kudritzki00} and the revisions by \cite{markova04}, we
determined the wind momenta and hence mass-loss rates at most grid
points.  Note, however, that the wind momentum-luminosity relationship
of eq. (3), which seems to hold for supergiants and O-type giants,
apparently breaks down in O-type dwarfs with luminosities $\log
(L/\Lsol) < 5.2$\,.  For these stars we adopted the relation $\log
D_{\rm mom} = 12 + 3 \times \log (L/\Lsol)$ to account for their much
weaker winds.  The same formula was applied to B-type giants and
dwarfs with luminosities $ \log (L/\Lsol) > 4.0$\,.  For lower
luminosities a very low value $\log D_{\rm mom} = 12$ was adopted to
simulate a very weak, unobservable, stellar wind.  Values of $\dot{M}$
and $v_{\infty}$ for solar metallicity are listed in the last two
columns of Table~\ref{tab:wmb_grid}.

Given that stellar winds are believed to be driven by the transfer of
photon momentum due to the absorption and scattering of radiation by
{\em metal\/} lines, it is expected that the wind properties will be
sensitive to the metallicity. To test this assertion observationally,
the properties of Galactic hot stars and those in the metal-poor Large
and Small Magellanic Clouds have been compared \citep{puls96}.  While
there is clear evidence for smaller terminal velocities in SMC stars
\citep{walborn95a}, further data are required to quantify the
dependence of the mass-loss rate on metallicity.  Nevertheless, there
is an obvious trend for smaller wind momenta in lower metallicity O
stars, and investigations are now underway to see whether this is also
the case for B and A supergiants.  In our models we adopted the best
theoretical estimates for the scaling of $v_{\infty}$ and $\dot{M}$
with metallicity, as given by \cite{leitherer92}, \cite{vink01}, and
\cite{kudritzki02,kudritzki03}:
\begin{equation}\label{eq:v_term_Z}
v_{\infty}=v_{\infty,\,\sol}\,{\left({Z \over \Zsol}\right)^{0.13}}
\quad{\rm and}
\end{equation}
\begin{equation}\label{eq:m_dot_Z}
\dot{M}=\dot{M}_{\sol}\,{\left({Z \over \Zsol}\right)^{0.69}} .
\end{equation}

Note that the hydrodynamic structure of the {\em WM-basic\/} models is
calculated by accounting for a radiative line force in the equation of
motion, parameterized by a `line force multiplier'. In order to
calculate models with pre-specified values of the mass-loss rate and
terminal velocity, it is necessary to estimate values of this
parameter.  To this end, we used the analytical approach by
\cite{kudritzki89} in an iterative algorithm to provide the {\em
  WM-basic\/} code with force multipliers that yield the appropriate
values of $v_{\infty}$ and $\dot{M}$ as the result of the solution to
the equation of motion.

When running the grid at the five metallicities (2, 1, 0.4, 0.2 and
0.05\zsol), we had the choice of using either (i)~{\em WM-basic's\/}
solar reference abundances scaled by a constant factor or (ii)~our own
choice of individual element abundances. We first consider the case of
modelling star-forming regions within our Galaxy of `solar'
metallicity. Two possible sets of element abundances are the solar
scale and the Orion nebula scale; we compare them in
Table~\ref{tab:orion_abundances}. The differences between the two,
particularly in some of the element ratios [e.g. (N/O), (Mg/O) and
(Si/O)] are still a source of concern; presumably they reflect
unresolved systematic errors in either, or both, sets of measurements.
In the present work, we adopted the Orion nebula values for elements
included in the compilation by \cite{esteban98}; column (3) in
Table~\ref{tab:orion_abundances} lists the corresponding correction
factors we applied to the default solar scale provided by {\em
  WM-basic\/}. For elements not included in the compilation by
\cite{esteban98} we used the default solar scale in {\em WM-basic\/}.

%%%%%%%%%%%%%%%%%%%%%%%%%%%%%%%%%%%%%%%%%%%%%%%%%%%%%%%%%%%%%%%%%%%%%%
% Increase the table counter by one, as the first table is long and 
% must go at the end.
\addtocounter{table}{1}

\begin{deluxetable*}{lccccc}
\tablewidth{275.42789pt}
\tablecaption{\textsc{Orion Abundance Scale}\label{tab:orion_abundances}}
\tablehead{
  \colhead{X}
& \colhead{12+log(X/H)$_{\rm Orion}$\tablenotemark{a}}
& \colhead{Correction \tablenotemark{b}}
& \colhead{12+log(X/H)$_{\rm \sol}$\tablenotemark{c}}
& \colhead{Ref \tablenotemark{d}}
& \colhead{[X/H]$_{\rm Orion}$\tablenotemark{e}} }
\startdata
C  &  8.49  & $-0.029$     &  8.44  & 1 & $+$0.05 \\
N  &  7.78  & $-0.139$     &  7.93  & 2 & $-$0.15 \\
O  &  8.72  & $-0.112$     &  8.74  & 2 & $-$0.02 \\
Ne &  7.89  & $-0.189$     &  8.00  & 2 & $-$0.11 \\
Mg &  7.43  & $-0.150$     &  7.58  & 3 & $-$0.15 \\
Si &  7.36  & $-0.184$     &  7.56  & 3 & $-$0.20 \\
S  &  7.17  & $-0.152$     &  7.20  & 3 & $-$0.03 \\
Cl &  5.33  & $-0.175$     &  5.28  & 3 & $+$0.05 \\
Ar &  6.49  & $+0.093$     &  6.40  & 3 & $+$0.09 \\
Fe &  7.48  & $-0.025$     &  7.50  & 3 & $-$0.02 \\
\enddata
\tablenotetext{a}{Orion nebula abundances from the compilation by
  \cite{esteban98}.}
\tablenotetext{b}{Logarithmic correction factors applied to the
  internal {\em WM-basic} solar scale, in order to obtain the Orion
  abundances in column (1).}
\tablenotetext{c}{Solar abundance scale.}
\tablenotetext{d}{References for the solar abundance scale (see below).}
\tablenotetext{e}{[X/H]$_{\rm Orion}$=log(X/H)$-$log(X/H)$_{\rm \sol}$.}  
\tablerefs{(1)~\cite{allende-prieto02} --- note that we adopt the C/O
  ratio from this source and apply it to our solar oxygen abundance;
  (2)~\cite{holweger01}; (3)~\cite{grevesse98}.}
\end{deluxetable*}
%%%%%%%%%%%%%%%%%%%%%%%%%%%%%%%%%%%%%%%%%%%%%%%%%%%%%%%%%%%%%%%%%%%%%%

Of course there is no reason {\it a priori} why the mix of different
elements should be the same as in the Orion nebula (or the Sun).
Element ratios can (and do) vary as a function of metallicity,
reflecting the past history of star formation of a galaxy. Departures
from solar ratios are plausible in high redshift star-forming
galaxies, given that these objects generally support much higher star
formation rates than our Galaxy did at any time in its past. Under
these circumstances, elements synthesized by long-lived stars,
primarily iron and other Fe-peak elements, may well be underabundant
relative to the products of massive stars (oxygen and other
$\alpha$-capture elements), as found by Pettini et al. (2002) in
MS1512--cB58.  While one of the advantages of our fully theoretical
approach to synthesizing the UV spectra of star-forming galaxies is
that we could in principle choose any mix of elements, for this first
attempt we decided to simply scale the Orion nebula abundance pattern
to different metallicities, keeping the relative element ratios fixed
at their values in the $Z = 1 \Zsol$ model listed in
Table~\ref{tab:orion_abundances}.  If the results of the work
presented here are sufficiently promising, it will be possible in
future to generate models with different abundance ratios to, for
example, determine the level of $\alpha$-element enhancement in the
galaxies under study.

\subsection{Integration of the Spectral Libraries into {\it Starburst99\/}}\label{subsec:starburst99}

Ours is the latest in a series of developments and improvements to
expand the capability of the spectral synthesis code {\em
  Starburst99\/} since it was first released in September 1998.  The
original version \citep[see][]{leitherer99} did allow some parameters
to be calculated at five metallicities (2, 1, 0.4, 0.2 and 0.05\zsol)
thanks to the inclusion of evolutionary tracks and model atmospheres
at these metallicities. However, it was only possible to synthesize
the emergent {\em ultraviolet spectrum\/} of a star-forming region of
solar metallicity because empirical stellar libraries were available
only for Milky Way O-type stars. The inclusion of spectra of Galactic
B stars in October 1999 \citep[][]{demello00} allowed the useful
spectral range of the synthesized spectra to be extended to 1850\,\AA\ 
where these cooler stars make a significant contribution to the
integrated light.  The next major development to the code was the
introduction in December 2000 of a library of O-type spectra assembled
from {\em HST} STIS observations of stars in the Large and Small
Magellanic Clouds \citep{leitherer01}, giving the user the option to
synthesize the UV spectrum of a population with a `hybrid' metallicity
of about 1/4 solar, albeit over the restricted wavelength range
1250--1600\,\AA.  Most recently, \cite{smith02} created a grid of 230
expanding non-LTE line-blanketed model atmospheres, at 2, 1, 0.4, 0.2
and 0.05\zsol, and incorporated them into {\em Starburst99\/} for the
purpose of calculating realistic ionizing fluxes from young stellar
populations.

\cite{smith02} utilized two sets of model atmospheres: the {\em
  WM-basic\/} code by \cite{pauldrach01} for O stars and the {\em
  CMFGEN\/} code by \cite{hillier98} and \cite{hillier99} for
Wolf-Rayet stars.  Thus, their O-star models were calculated with the
same code used in the present study. Since \cite{smith02} focused on
the emergent ionizing continuum below 912\,\AA\ and not on the line
spectrum at longer wavelengths, the spectral resolution of their
models is not sufficient for the absorption line studies that are the
subject of this paper.  However, the continuous spectra predicted by
\cite{smith02} and by us are identical---unless Wolf-Rayet stars were
important contributors to the continuum from the integrated stellar
population, which is usually not the case.

%%%%%%%%%%%%%%%%%%%%%%%%%%%%%%%%%%%%%%%%%%%%%%%%%%%%%%%%%%%%%%%%%%%%%%
\begin{figure*}
\begin{center}
\includegraphics[angle=270,width=0.9\textwidth]{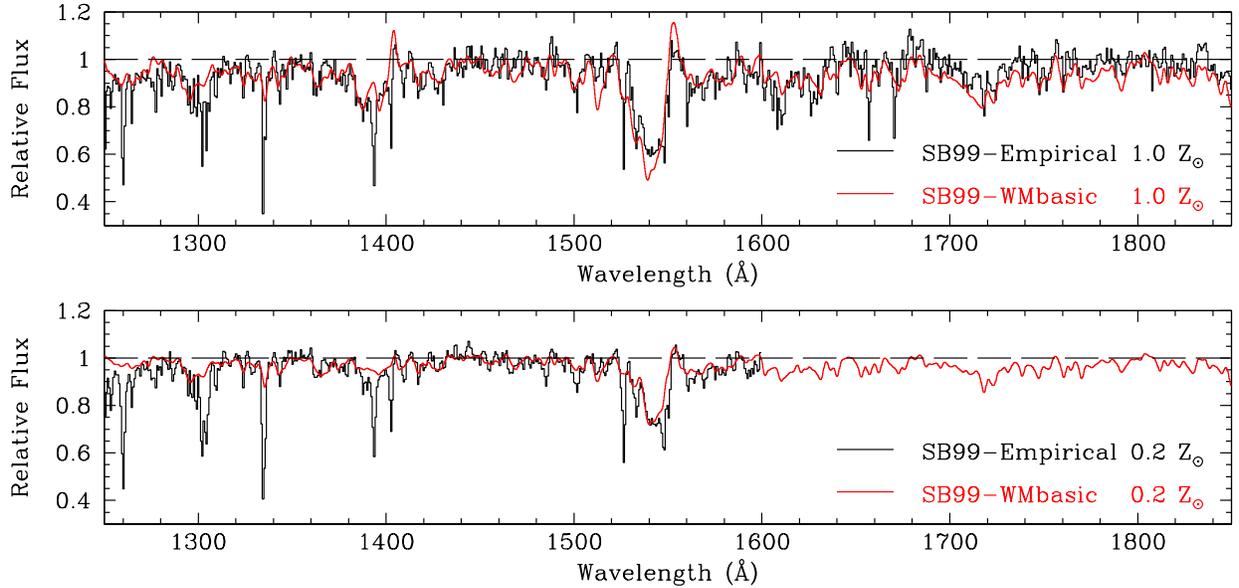}
\caption{Comparison of two {\em Starburst99 + WM-basic\/} 
spectra ({\em red\/}) with `conventional' {\em Starburst99\/} output
spectra generated with empirical libraries of Galactic and Magellanic 
Clouds OB stars ({\em black\/}). 
The models shown are for continuous star formation (sampled 
at an age of 100\,Myr when the UV spectral features have fully 
stabilized) with a Salpeter slope of the IMF between 1 and
100\msol\ and metallicities as indicated. The integrated spectra 
synthesized with empirical stellar libraries include narrow
interstellar absorption lines which are not present in the fully
theoretical spectra. The libraries of Magellanic Cloud stars used
by {\em Starburst99\/} only extend to 1600\,\AA; this accounts for the
reduced range of the `SB99-Empirical' spectrum in the lower panel.
\label{fig:emp_theor_comp}}
\end{center}
\end{figure*}
%%%%%%%%%%%%%%%%%%%%%%%%%%%%%%%%%%%%%%%%%%%%%%%%%%%%%%%%%%%%%%%%%%%%%%

For the present work, we built a non-public version of {\it
  Starburst99\/} in which the empirical UV library of {\it IUE\/} and
{\it HST\/} spectra was replaced by the grid of theoretical spectra
generated with {\em WM-basic\/} as described in
\S\ref{subsec:wmbasic}.  Most of the implementation followed the same
steps as the standard version of {\it Starburst99\/}, but with two
significant differences, as we now explain.

First, the theoretical spectra are not characterized by spectral types
but by their effective temperatures and gravities.  Therefore, we can
immediately link them to the positions in the H-R diagram predicted by
the stellar evolutionary tracks.  This is a major advantage over
empirical libraries because it avoids the intermediate step of having
to adopt a spectral type vs. effective temperature calibration.
Recent theoretical and empirical work
\citep{crowther02,bianchi02,martins02} has shown that the effective
temperature scale of O stars needs to be revised downward from
previous calibrations. Such a revision affects spectral synthesis
models based on the empirical libraries but not those generated
theoretically.

Second, model atmospheres predict both the idealized line-free
continuum, and the apparent continuum which includes line blanketing.
Therefore we can link the theoretical spectra to the evolutionary
models without having to normalize them first.  Empirical UV spectra
are always affected by an unknown amount of the interstellar
reddening. As a result, they must be rectified prior to adding them to
a library. Then, the rectified spectra are recalibrated to flux units
by fitting a theoretical low-resolution continuum at each grid point
in the H-R diagram.  This process introduces additional uncertainties
which are avoided when a purely theoretical method is employed.

We performed extensive tests with the synthesized spectra.  As
discussed below (\S2.3), the computed photospheric features agree well
with those generated with empirical stellar libraries.  In addition,
the stellar wind lines of \ion{C}{4}~$\lambda\lambda 1548, 1550$ and
\ion{Si}{4}~$\lambda\lambda 1393, 1402$ show reasonable agreement.
However, we believe that a more detailed study of all wind lines is
warranted; since these lines are not utilized in the present analysis,
their discussion is deferred to a future paper.  Once work on the wind
lines is completed, we will release an updated version of the {\em
  Starburst99\/} code which includes the new theoretical libraries.

With our new, fully theoretical, code we can now model the spectra of
star-forming regions as a function of time, for either an
instantaneous burst or a continuous star formation law and with a
choice of the initial mass function (IMF) slope and mass range, for
metallicities of 2, 1, 0.4, 0.2 or 0.05\zsol.

\subsection{Comparison with the {\it Starburst99\/} Empirical Libraries}\label{subsec:comp_emp_spec}

Before using our new code to investigate the effects of metallicity on
the integrated stellar spectrum, we performed a `sanity check' by
comparing the outputs of two of our {\em Starburst99 + WM-basic\/}
models with the corresponding model spectra generated in the
conventional manner, that is using empirical libraries of stellar UV
spectra.  Since the latter are available for two metallicities,
corresponding to Galactic and Magellanic Cloud stars respectively, we
compared them with the fully theoretical spectra generated with $Z =
1$\zsol\ and 0.2\zsol.\footnote{The empirical libraries for Magellanic
  Cloud stars include stars from both the Large and Small Magellanic
  Clouds, giving a `hybrid' metallicity $Z_{\rm MC} \approx 0.25
  \Zsol$.  We compare them with our closest theoretical libraries, for
  $Z = 0.2 \Zsol$.  In both the empirical and theoretical case, the
  evolutionary tracks used by {\em Starburst99\/} are for $Z = 0.2
  \Zsol$.}  The comparison is illustrated in
Figure~\ref{fig:emp_theor_comp} for our `standard' case of a
continuous star formation model with a Salpeter IMF between 1 and
100\msol\ (see \S\ref{sec:synthetic_spectra}).  The spectra generated
with the {\em WM-basic\/} libraries were smoothed and normalised
according to the procedures described in \S\ref{subsec:continuum_fit}.

It is important to realise, when comparing the two sets of spectra,
that they differ in their interstellar absorption. The empirical
stellar spectra, obtained from observations of real stars, include the
most prominent UV interstellar absorption lines which can be strong,
particularly along sight-lines to the Large Magellanic Cloud.  The
{\em WM-basic\/} spectra, on the other hand, do not suffer from this
contamination and are purely stellar.  When this difference is taken
into account, Figure~\ref{fig:emp_theor_comp} shows that there is a
reasonably good agreement between the spectra produced by the
empirical and theoretical libraries at both metallicities. Some of the
discrepancies, such as those at wavelengths greater than 1700\,\AA\ in
the solar metallicity models, appear to arise from the continuum
normalization rather than from intrinsic differences.  Of particular
interest for the present work are two blends of photospheric
absorption lines, at 1360--1380\,\AA\ (which also includes the weak
stellar wind line \ion{O}{5}~$\lambda 1371$) and 1415--1435\,\AA,
which \cite{leitherer01} showed to be sensitive to metallicity.
Inspection of Figure~\ref{fig:emp_theor_comp} reveals a fairly good
correspondence in these wavelength regions.

\section{SYNTHETIC UV SPECTRA AT METALLICITIES
  $Z=0.05-2\Zsol$}\label{sec:synthetic_spectra}

\subsection{Computation of the {\it Starburst99\/} Models}\label{subsec:computation_sb99}

Having confirmed that our new version of {\em Starburst99\/} produces
results that are consistent with those generated by the empirical
libraries, we can now turn to its application to the study of high
redshift star-forming galaxies. In order to do so, it is necessary
first of all to specify the mode of star formation. {\em
  Starburst99\/} considers two limiting cases, either an instantaneous
(i.e. unresolved in time) burst in which all the stars form at time $t
= 0$ and the stellar population then evolves passively, or continuous
star formation at a constant rate.  This distinction is important for
our purposes because in the burst scenario the integrated spectrum
changes rapidly with time, on timescales of a few Myr, reflecting the
short lifetimes of the O and B type stars. In this limit, the emergent
spectrum is more sensitive to the age than the metallicity of the
stellar population.  On the other hand, in the constant star formation
limit, a quasi-equilibrium is reached within a few tens of Myr, after
which there is little spectral variation with time
\citep[e.g.][]{leitherer99,leitherer01}.  Clearly, such a situation is
much more favourable to the use of stellar features as metallicity
indicators.

%%%%%%%%%%%%%%%%%%%%%%%%%%%%%%%%%%%%%%%%%%%%%%%%%%%%%%%%%%%%%%%%%%%%%%
\begin{figure*}
\begin{center}
\includegraphics[angle=270,width=0.85\textwidth]{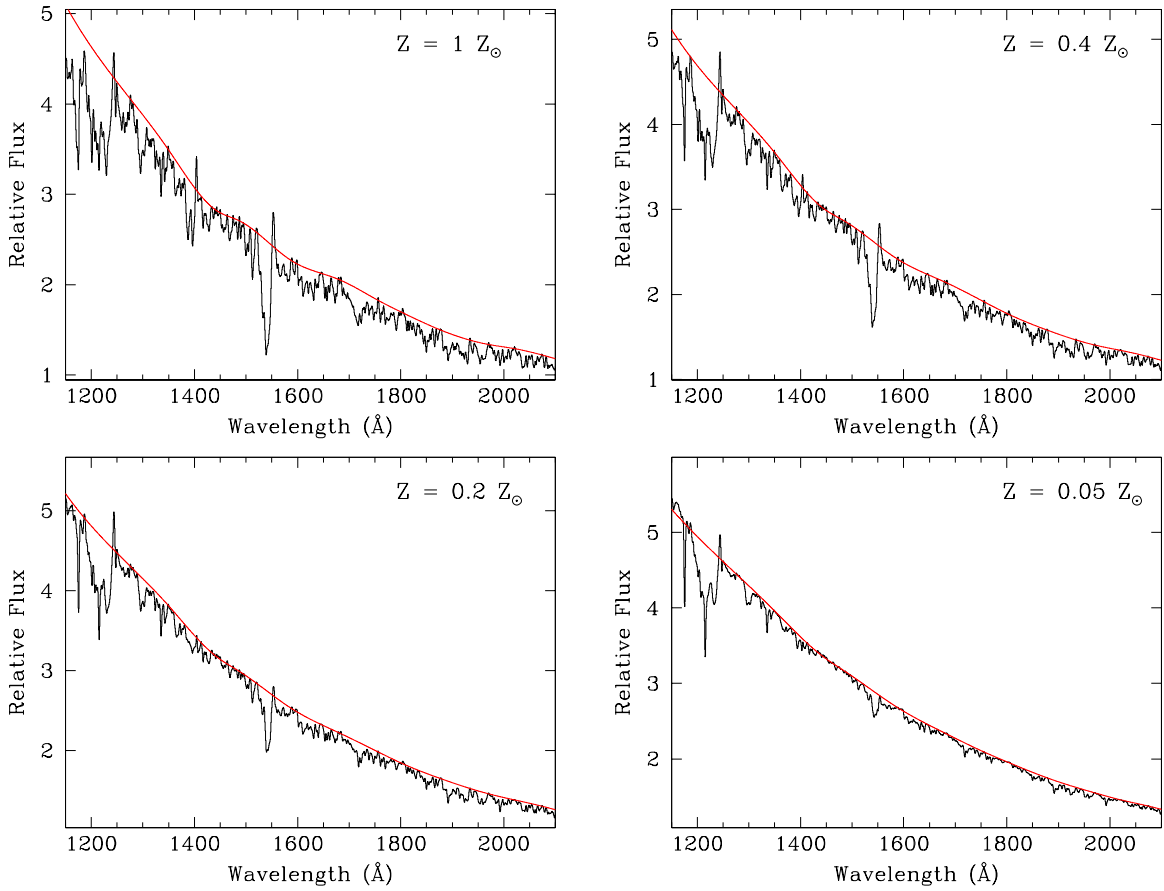}
\caption{Examples of the continuum fitting process, showing the spline
curves fitted to the `pseudo-continuum' points
listed in Table~\ref{tab:continuum_windows}.
The model spectra shown are for our standard case of continuous
star formation (at age 100\,Myr) with a Salpeter slope of the 
IMF between 1 and 100\msol\ and metallicities as indicated.
\label{fig:spline_fit}}
\end{center}
\end{figure*}
%%%%%%%%%%%%%%%%%%%%%%%%%%%%%%%%%%%%%%%%%%%%%%%%%%%%%%%%%%%%%%%%%%%%%%

%%%%%%%%%%%%%%%%%%%%%%%%%%%%%%%%%%%%%%%%%%%%%%%%%%%%%%%%%%%%%%%%%%%%%%
\begin{deluxetable}{ccc}
\tablewidth{113.82433pt}
\tablecaption{\textsc{Continuum Points}\label{tab:continuum_windows}}
\tablehead{
  \colhead{Window}
& \colhead{$\lambda_{\rm min}$ (\AA)}
& \colhead{$\lambda_{\rm max}$ (\AA)}
}
\startdata
\phn1 & 1274.5 & 1278.0 \\
\phn2 & 1348.0 & 1351.0 \\
\phn3 & 1439.5 & 1444.5 \\
\phn4 & 1485.5 & 1488.5 \\
\phn5 & 1586.0 & 1590.5 \\
\phn6 & 1678.0 & 1681.0 \\
\phn7 & 1754.5 & 1758.0 \\ 
\phn8 & 1800.0 & 1804.0 \\ 
\phn9 & 1875.0 & 1880.0 \\ 
   10 & 1968.0 & 1972.0 \\ 
   11 & 2018.0 & 2023.0 \\
   12 & 2072.0 & 2075.0 \\
   13 & 2110.0 & 2113.0 \\
\enddata             
\end{deluxetable}    
%%%%%%%%%%%%%%%%%%%%%%%%%%%%%%%%%%%%%%%%%%%%%%%%%%%%%%%%%%%%%%%%%%%%%%

Both modes of star formation considered by {\em Starburst99\/} are
idealised approximations to the real progress of star formation in
galaxies, which is likely to be a series of bursts.  However, when
analysing the integrated UV stellar output from a {\em whole} galaxy,
the continuous star formation mode seems to be the better description,
at least for most Lyman break galaxies, as we now explain.  If a
single burst followed by passive evolution were a valid approximation,
we would expect to see large numbers of post-starburst galaxies in
which the most massive stars have already died.  While still bright in
the UV continuum thanks to the longer-lived B stars, such galaxies
would lack the nebular emission lines whose strength is determined
primarily by the hottest stars present. In other words, while the
optical emission lines would fade away after $\sim 10$\,Myr, the
galaxies would remain UV bright, and still meet the colour criteria
tuned to the photometric selection of star-forming galaxies at
redshifts $z > 1$ \citep{adelberger04}, for several tens of Myr.
Galaxies with these characteristics seem to be the exception in the
surveys by Steidel and collaborators \citep{pettini01,erb03}.

In any case, causality arguments alone argue for star formation
timescales greater than $30-40$\,Myr.  This seems to be the
characteristic dynamical timescale $\tau_{\rm dyn} = 2r_{1/2}/\sigma$
of Lyman break galaxies at $z \simeq 3$, which typically have
half-light radii $r_{1/2} \simeq 1-2 h^{-1}$\,kpc and velocity
dispersions $\sigma \simeq 75 \pm 25$\kms\ \citep{pettini01}.  Similar
values of $\tau_{\rm dyn}$ apply to photometrically selected
star-forming galaxies at $z \simeq 1.5 - 2.5$ [the `BX' and `BM'
galaxies surveyed by \cite{steidel04} and whose kinematics have been
studied by \cite{erb03,erb04}].  On the basis of these considerations,
we only consider continuous star formation models in the present work.

The second assumption which underlies all stellar population modelling
is the shape and the mass limits of the IMF, which is usually
approximated by a power law of the form
\begin{equation}\label{eq:imf}
\phi(m)=\frac{{\rm d}n}{{\rm d}m}=Cm^{-\alpha},
\end{equation}
(or a combination of such power laws), where $n$ is the number of
stars, $m$ is their mass, $C$ is a normalization constant and $\alpha$
is the IMF slope.  The slope $\alpha=2.35$ at the upper mass end
originally proposed by \cite{salpeter55} has stood the test of time
\citep{kennicutt98} and seems to apply to high redshift galaxies too
\citep{pettini00,steidel04}.  Accordingly, in all the modelling
presented here we adopted a universal IMF with a Salpeter slope
between 1 and 100\msol\ (these being the usual limits used in {\em
  Starburst99\/}).  However, in \S\ref{sec:imf_comparison} we also
consider the consequences of adopting different parameters for the
IMF.

\subsection{Continuum Fitting and  Resolution}\label{subsec:continuum_fit}

With the assumption of a continuous star formation mode, with a
Salpeter IMF between 1 and 100\msol\ and a constant star formation
rate, we used our new version of {\em Starburst99\/} to compute
synthetic spectra, at time intervals of 5\,Myr from 5 to 150\,Myr, for
each of the five metallicities considered.  Before analysing these
spectra, however, it is necessary to perform a couple of processing
steps, as we now explain.

In order to measure any spectral features (whether wind, photospheric,
or interstellar lines) it is necessary to first normalize the spectra.
This process removes the shape of the underlying continuum by fitting
an appropriate function to regions deemed to be free of
emission/absorption features and then dividing the spectrum by this
function.  Second, a comparison between observed and synthesized
spectra is best carried out at comparable spectral resolution so that,
in general, it is necessary to smooth one spectrum to match the
resolution of the other.  The important point to be aware of is that
continuum fitting and spectral resolution are intimately related.
This is because in reality there are very few, if any, wavelength
intervals in the UV spectrum of a young stellar population that are
truly free of photospheric absorption lines. Thus, smoothing the
spectrum invariably results in the depression of even relatively
line-free regions and produces an apparent `pseudo-continuum' which is
lower than the real continuum level, now no longer discernible.

The primary motivation of this work is to generate a set of fully
theoretical spectra which can be compared with those of star-forming
galaxies at high redshifts. The largest samples of such galaxies are
from the spectroscopic surveys by Steidel and collaborators
\citep{steidel03, steidel04} conducted with the Low Resolution Imaging
Spectrograph (LRIS) on the Keck telescopes. The resolution of this
data set, FWHM\,$\simeq 2.5$\,\AA\ in the rest frame, or $\simeq
500$\kms\ at 1500\,\AA, is the major source of line broadening, and
dominates over the broadening from internal stellar processes, stellar
rotation, and the velocity dispersion of the stars within a galaxy.
We therefore simply convolved the synthetic spectra generated by {\em
  Starburst99 + WM-basic\/} with a Gaussian of FWHM\,$= 2.5$\,\AA.  In
order to define the continuum level, we used both our model spectra
and actual spectra of Lyman break galaxies to select
`pseudo-continuum' points which are listed in
Table~\ref{tab:continuum_windows}. For each model spectrum, we fitted
a spline curve through the mean flux in each of these narrow windows
and divided by this fit to produce a normalized spectrum.
Figure~\ref{fig:spline_fit} shows some examples of the spline curves
fitted to the models.

%%%%%%%%%%%%%%%%%%%%%%%%%%%%%%%%%%%%%%%%%%%%%%%%%%%%%%%%%%%%%%%%%%%%%%
\begin{figure*}
\center
\includegraphics[angle=270,width=0.8\textwidth]{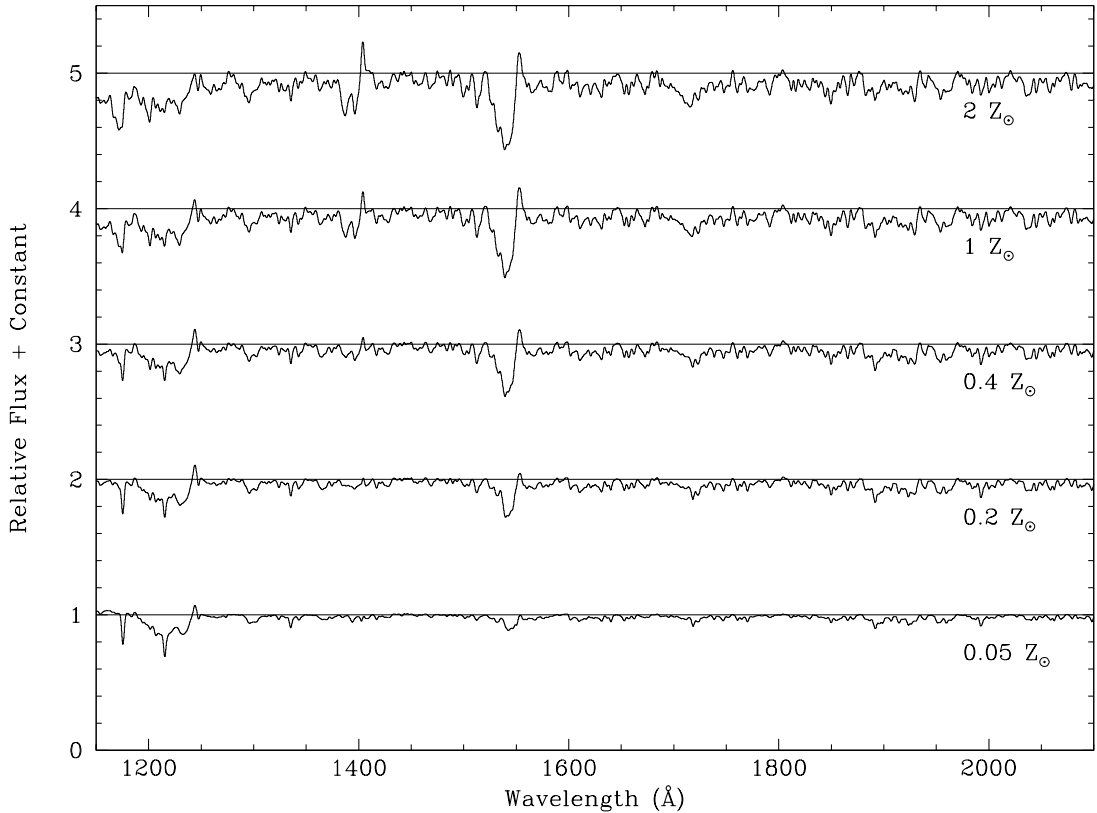}
\caption{Fully synthetic UV spectra of star-forming regions of
metallicities 2--0.05\zsol. All the models shown are for our standard case 
of continuous star formation (at age 100\,Myr) with a Salpeter slope of the 
IMF between 1 and 100\msol.
Each spectrum has been convolved with a Gaussian of 
FWHM\,$= 2.5$\,\AA\ and divided by a spline fit to pseudo-continuum
points (see Figure~\ref{fig:spline_fit}).
\label{fig:synthetic_spectra}}
\end{figure*}
%%%%%%%%%%%%%%%%%%%%%%%%%%%%%%%%%%%%%%%%%%%%%%%%%%%%%%%%%%%%%%%%%%%%%%

\subsection{The Sensitivity of the Stellar Spectrum to Metallicity}\label{subsec:sensitivity_Z}

Figure~\ref{fig:synthetic_spectra} presents the smoothed and
normalised spectra of the 2, 1, 0.4, 0.2 and 0.05\zsol\ models,
100\,Myr after the onset of star formation (by this time the stellar
features in the composite UV spectrum have fully stabilised and show
no residual time variation).  Examination of the figure shows that the
spectral appearance {\em is\/} strongly altered by the presence of
metals. The most obvious change, as the metallicity decreases, is in
the declining strengths of both the absorption and emission components
of the stellar wind lines, especially \ion{C}{4}~$\lambda \lambda
1548, 1550$.  Such a marked dependence on metallicity has recently
been reported in all the strongest stellar wind lines by \cite{keel04}
in their study of star-forming regions in nearby galaxies at far-UV
wavelengths with the {\em Far-Ultraviolet Spectroscopic Explorer\/}.
While it may be tempting to calibrate these features in terms of
metallicity, unfortunately (for the present purposes)
\ion{C}{4}~$\lambda \lambda 1548, 1550$ is normally blended with
strong interstellar absorption in the spectra of high redshift
galaxies \citep[see, for example, Fig.~4 of ][]{pettini03_lanzarote}.
Furthermore, the strengths and profiles of these lines respond not
only to metallicity, but also to the slope and upper mass cutoff of
the IMF (\S\ref{sec:imf_comparison}) and to time dependent dust
obscuration \citep{leitherer02}.  Although there have been attempts to
interpret the \ion{C}{4} (and \ion{Si}{4}~$\lambda \lambda 1394,
1403$) lines in terms of metallicity \citep[e.g.][]{mehlert02}, a
proper treatment requires careful deblending of the different
components (photospheric, wind, and interstellar) which contribute to
these composite spectral features.  These complicating factors are
much less important for the numerous blends of photospheric absorption
lines which make up most of the UV spectrum of star-forming regions
and which can also be seen from Figure~\ref{fig:synthetic_spectra} to
scale with metallicity.

Following the inclusion of the low metallicity library of {\em
  empirical\/} stellar spectra into Starburst99, \cite{leitherer01}
investigated the influence of metallicity upon the spectral appearance
of a star-forming region. They drew attention to two blends of lines
near 1370\,\AA\ and 1425\,\AA\ (which they attributed to
\ion{O}{5}~$\lambda 1371$ and \ion{Fe}{5}~$\lambda \lambda 1360-1380$,
and to \ion{Si}{3}~$\lambda 1417$, \ion{C}{3}~$\lambda 1427$, and
\ion{Fe}{5}~$\lambda 1430$ respectively) and showed that the
equivalent widths of these blends decrease by factors of 2--4 as the
metallicity of the stellar population drops from solar to $\sim 1/4$
solar.

We can verify and extend these results with our {\em Starburst99 +
  WM-basic\/} models.  Figures~\ref{fig:plot_ew_indices_1370} and
\ref{fig:plot_ew_indices_1425} show the equivalent widths we measure
for the `1370' and the `1425' line indices [integrated, as prescribed
by \cite{leitherer01}, over the wavelength ranges 1360--1380\,\AA\ and
1415--1435\,\AA\ respectively] for different metallicities and ages.
These measurements confirm that the lines indices increase
monotonically with metallicity over the range $Z = 0.05-2 \Zsol$
considered here.  Equally important is the fact that the indices are
stable after $\sim 50$\,Myr from the onset of star formation, at which
point a quasi-equilibrium is reached with approximately constant
relative numbers of each type of star, even though individual stars
continuously form and die.  Dependence on metallicity and stability
with time are both necessary conditions to avoid a degeneracy between
age and metallicity which would compromise the usefulness of a line
index.

Although our model spectra support the initial suggestion by
\cite{leitherer01} that the 1370 and 1425 indices may be useful
metallicity indicators, these two spectral features are far from ideal
for this purpose. First, they are weak, with equivalent widths of only
$\lesssim 1.5$\,\AA\ at $Z \simeq \Zsol$; therefore their use as
metallicity diagnostics requires data of high S/N.  A second
complication is the fact that these lines are blends of transitions
from more than one element and therefore only give some (poorly
defined) `average' metallicity. It would not be possible, for example,
to use them to investigate departures of different elements from their
solar ratios.  Finally, their far-UV wavelengths make them difficult
to measure from the ground at redshifts lower than $z \sim 1.7$,
thereby excluding the large numbers of star-forming galaxies now being
discovered at these intermediate redshifts \citep{steidel04,
  abraham04}.

\subsection{The Region 1935--2020\,\AA}

Aware of these restrictions, we considered other UV stellar features
which do not suffer from the same problems. The most suitable
candidate we have found is in the range $\sim$1900--2050\,\AA.  The
broad blend of lines in this region was identified thirty years ago
with the {\em S2/68 Ultraviolet Sky-Survey Telescope\/} on board the
TD1 satellite which provided one of the first views of the sky at
ultraviolet wavelengths. From spectrophotometric measurements of
early-type stars at a resolution of 35\,\AA, \cite{thompson74}
reported the presence of a broad absorption region between 1900\,\AA\ 
and 1940\,\AA, which they attributed to line blocking by numerous
photospheric \ion{Fe}{3} transitions \citep{elst67}.  A more extensive
study by \cite{swings76} confirmed that many of the absorption
features between 1800\,\AA\ and 2200\,\AA\ are indeed \ion{Fe}{3}
transitions.  These authors also showed that the strength of the
absorption is greater in B-stars of earlier types and higher
luminosity classes.  \cite{nandy81} observed a similar trend for B
stars in the Large Magellanic Cloud and noted that the absorption is
much weaker in O stars. In a conference proceedings, \cite{morgan86}
addressed the question of whether the \ion{Fe}{3} feature is sensitive
to metallicity by studying stars in the metal-poor Small Magellanic
Cloud. Their results showed that, on average, the B star absorption is
weaker at lower metallicities.

In a study of the ultraviolet properties of nearby starbursts,
\cite{heckman98} used {\it IUE} spectra of $\sim 20$ galaxies to
create composite spectra with mean metallicities of $\sim 0.16 \Zsol$
and $\sim 1.1 \Zsol$. Their work showed that not only is the
\ion{Fe}{3} absorption clearly visible in the spectra of starburst
galaxies but also confirmed that this feature is indeed stronger in
higher metallicity environments.  In addition to these encouraging
preliminary indications, the absorption at $\sim 1900-2050$\,\AA\ 
overcomes many of the difficulties discussed above for the 1370 and
1425 indices.  It is broader and stronger, it should be mainly
sensitive to the iron abundance, and is accessible from the ground for
galaxies down to $z \sim 1$.  It therefore merits a closer look.

%%%%%%%%%%%%%%%%%%%%%%%%%%%%%%%%%%%%%%%%%%%%%%%%%%%%%%%%%%%%%%%%%%%%%%
\begin{figure}
\includegraphics[width=\columnwidth]{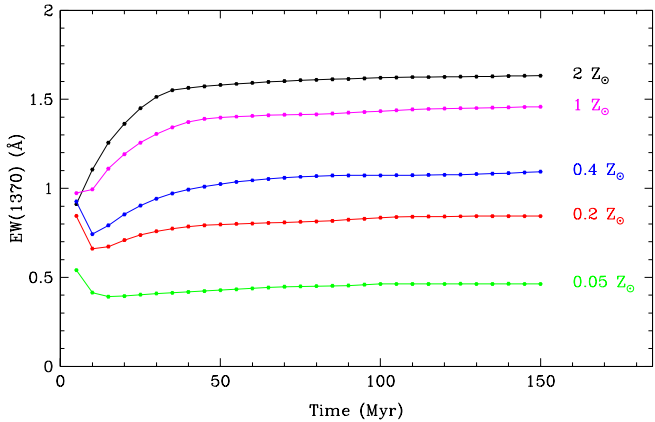}
\caption{Variation of the 1370 line index with time and
metallicity. \label{fig:plot_ew_indices_1370}}
\end{figure}
%%%%%%%%%%%%%%%%%%%%%%%%%%%%%%%%%%%%%%%%%%%%%%%%%%%%%%%%%%%%%%%%%%%%%%

%%%%%%%%%%%%%%%%%%%%%%%%%%%%%%%%%%%%%%%%%%%%%%%%%%%%%%%%%%%%%%%%%%%%%%
\begin{figure}
\includegraphics[width=\columnwidth]{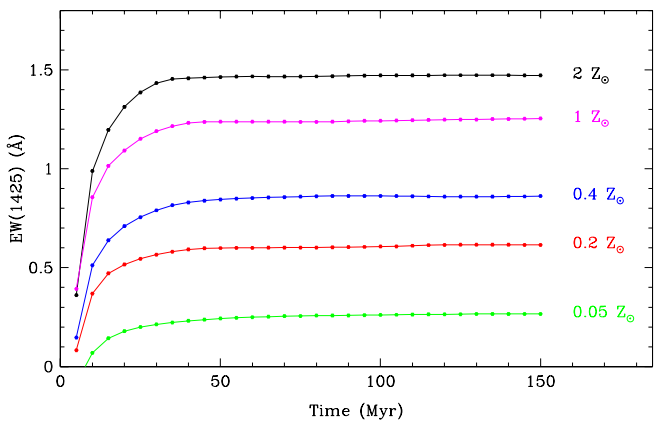}
\caption{Variation of the 1425 line index with time and
metallicity. \label{fig:plot_ew_indices_1425}}
\end{figure}
%%%%%%%%%%%%%%%%%%%%%%%%%%%%%%%%%%%%%%%%%%%%%%%%%%%%%%%%%%%%%%%%%%%%%%

%%%%%%%%%%%%%%%%%%%%%%%%%%%%%%%%%%%%%%%%%%%%%%%%%%%%%%%%%%%%%%%%%%%%%%
\begin{figure}
\includegraphics[width=\columnwidth]{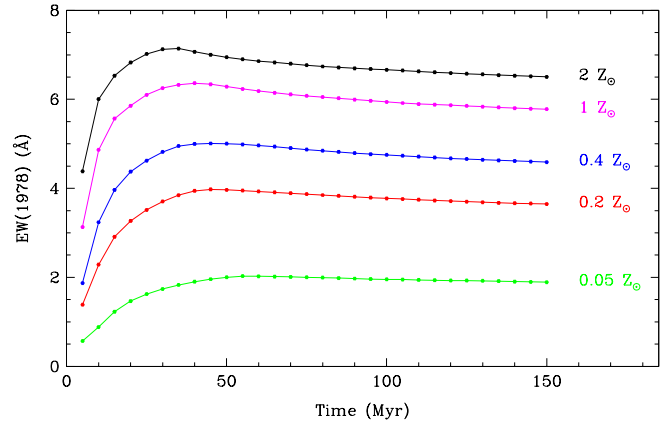}
\caption{Variation of the 1978 line index with time and
metallicity. \label{fig:plot_ew_indices_1978}}
\end{figure}
%%%%%%%%%%%%%%%%%%%%%%%%%%%%%%%%%%%%%%%%%%%%%%%%%%%%%%%%%%%%%%%%%%%%%%

Until now, it has not been possible to model this region of the
spectrum, {\em at any metallicity\/}, because the empirical stellar
libraries used by {\em Starburst99} do not extend to these mid-UV
wavelengths.  This is not a problem, however, for {\em WM-basic\/},
and the output from our runs was deliberately chosen to reach beyond
2000\,\AA\ (see Figure~\ref{fig:synthetic_spectra}).  The spectra
shown in Figure~\ref{fig:synthetic_spectra}, which are the first to
model the integrated spectrum of a star-forming region at mid-UV
wavelengths, indeed confirm that the strength of the \ion{Fe}{3} blend
increases with increasing metallicity.

To assess its variation with both time and metallicity we have defined
a new index, based upon the equivalent width measured between two
appropriate wavelengths. In choosing these wavelengths, we took into
consideration not only our synthetic models but also the observed
spectra of Lyman break galaxies, so as to avoid contamination with
other spectral features.  Interstellar absorption lines are
fortunately not a problem, provided we avoid \ion{Al}{3}~$\lambda
\lambda 1855, 1863$ which can be strong (the much weaker
\ion{Zn}{2}+\ion{Mg}{1} blend at $\lambda 2026$ is less of a concern).
Potentially more serious is the nebular emission line
\ion{C}{3}]~$\lambda 1909$ which is clearly seen in the composite
spectrum of $z \simeq 3$ galaxies generated by \cite{shapley03},
although its strength can apparently vary substantially between
different galaxies.

For these reasons, we decided to restrict the definition of our new
metallicity index to the equivalent width measured between 1935\,\AA\ 
and 2020\,\AA. This defines the `1978' index.  In our synthetic
spectra, these two integration limits appear to be at wavelengths
where the stellar pseudo-continuum is recovered.
Figure~\ref{fig:plot_ew_indices_1978} shows how the index varies with
time and metallicity. As was the case for the 1370 and 1425 indices,
the 1978 index increases monotonically with metallicity and reaches a
steady-state value after $\sim 50$\,Myr, thus satisfying both
requirements for a useful metallicity indicator.  It can also be
appreciated, by comparison of
Figures~\ref{fig:plot_ew_indices_1370}--\ref{fig:plot_ew_indices_1978},
that at $Z = \Zsol$ EW(1978) is $\sim 4$ times greater than EW(1370)
and EW(1425).

\section{COMPARISON WITH THE UV SPECTRA OF HIGH REDSHIFT
  GALAXIES}\label{sec:lbg_comparison}

%%%%%%%%%%%%%%%%%%%%%%%%%%%%%%%%%%%%%%%%%%%%%%%%%%%%%%%%%%%%%%%%%%%%%%
\begin{deluxetable*}{lcccccccccc}
\tablewidth{467.54414pt}
\tablecaption{\textsc{Observations of the Rest-Frame UV Spectra of
    High Redshift Galaxies}\tablenotemark{a}\label{tab:lbg_spectra}}
\tablehead{
  \colhead{Object}
& \colhead{$z_{\rm sys}$\tablenotemark{b}}
& \colhead{${\cal R}$}
& \colhead{Instrument\tablenotemark{c}}
& \colhead{Grating/Grism}
& \colhead{Exposure time}
& \colhead{$\Delta \lambda_{\rm obs}$}
& \colhead{$\Delta \lambda_{\rm rest}$}
& \colhead{Resolution\tablenotemark{d}}
& \colhead{SNR\tablenotemark{e}}
\\
  \colhead{}
& \colhead{}
& \colhead{}
& \colhead{}
& \colhead{(grooves mm$^{-1}$)}
& \colhead{(s)}
& \colhead{(\AA)}
& \colhead{(\AA)}
& \colhead{(\AA)}
& \colhead{}
}
\startdata
Q1307--BM1163  & 1.411\tablenotemark{f}\phn & 21.78 & LRIS B & \phn400 & \phn5400 & 3005 --\phn7105 & 1245 -- 2945 & 2.0\phn   & 37\\

\\
MS1512--cB58   & 2.7276\tablenotemark{g}      & 20.41 & ESI    & \phn175 &    16000 & 3860 --10500    & 1040 -- 2815 & \tablenotemark{\phm{h}}0.37\tablenotemark{h}      & 43\\
               &                              &       & LRIS R & \phn900 &    11400 & 4300 --\phn6020 & 1155 -- 1615 & 0.80      & 75\\
               &                              &       & LRIS R &    1200 & \phn3600 & 5875 --\phn7185 & 1580 -- 1925 & 0.56      & 24\\
\enddata

\tablenotetext{a}{All the observations were obtained with the Keck
  telescopes.}
\tablenotetext{b}{Systemic redshift.}
\tablenotetext{c}{B and R refer to the blue and red arms respectively
  of the LRIS spectrograph.}
\tablenotetext{d}{Spectral resolution at rest frame UV wavelengths.}
\tablenotetext{e}{Typical signal-to-noise ratio per resolution element.}
\tablenotetext{f}{From rest-frame optical nebular emission lines.}
\tablenotetext{g}{From the stellar photospheric C{\footnotesize ~III}~$\lambda
  2297$ line.}
\tablenotetext{h}{The resolving power of the ESI spectrum is 
$R = \lambda/\Delta \lambda = 5200$, which corresponds to 0.37\,\AA\ 
at a wavelength of 1900\,\AA.}
\end{deluxetable*}
%%%%%%%%%%%%%%%%%%%%%%%%%%%%%%%%%%%%%%%%%%%%%%%%%%%%%%%%%%%%%%%%%%%%%%

To summarize our work so far, we have computed theoretical spectra of
OB stars using the stellar wind code {\em WM-basic\/} and integrated
them into the spectral synthesis code {\em Starburst99\/}.  With this
combination, we have generated fully theoretical composite UV spectra
of star-forming regions for a range of metallicities, from twice solar
to 1/20 solar, under the simplest assumption of continuous star
formation with a Salpeter IMF.  We have analysed three blends of
photospheric stellar lines, at 1370, 1425, and 1978\,\AA, and found
them to be potentially useful metallicity indicators, in that their
equivalent widths increase monotonically with metallicity and change
little with time after $\sim 50$\,Myr (in the idealised approximation
of continuous star formation at a constant rate).  The 1978 index,
developed here for the first time, appears particularly promising in
that its equivalent width is larger than those of the other two, it is
free of contaminating interstellar lines, and is thought to be due
primarily to Fe lines.

Of course, the validity of these photospheric UV indices must be
tested against other established metallicity measures before they can
be applied with confidence to the study of high redshift galaxies.  At
present such a comparison is only possible in two cases, the galaxies
MS1512--cB58 and Q1307-BM1163.  These galaxies are unusually bright,
either because they are intrinsically luminous (Q1307-BM1163) or
gravitationally lensed (MS1512--cB58), and have been observed over a
range of wavelengths at high S/N, allowing measurements of element
abundances mostly from interstellar absorption and emission lines.
While the preliminary indications from these tests are encouraging, as
discussed below, clearly a larger sample of objects suitable for such
a cross-check must be assembled for a comprehensive assessment of the
method we have developed. We now consider each galaxy in turn;
Table~\ref{tab:lbg_spectra} gives details of observations of their
rest-frame UV spectra with the Keck telescopes.  The models shown in
the comparisons are all for our standard case of continuous star
formation with a Salpeter IMF at age 100\,Myr.  These models are
available in full from Table~\ref{tab:model_spectra}.

%%%%%%%%%%%%%%%%%%%%%%%%%%%%%%%%%%%%%%%%%%%%%%%%%%%%%%%%%%%%%%%%%%%%%%
\begin{figure}[b]
\includegraphics[width=\columnwidth]{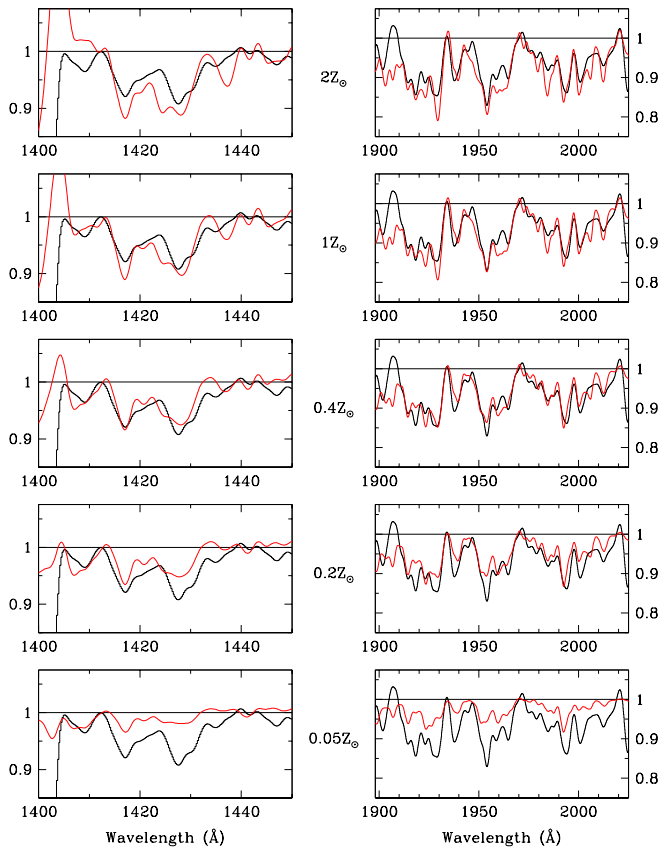}
\caption{Comparison of the observed spectrum of the 
$z = 2.7276$ Lyman break galaxy
MS1512--cB58 ({\em black}) with fully synthetic spectra
produced by the {\em Starburst99 + WM-basic\/}
combination ({\em red}) for five different
metallicities, from twice solar to 1/20 of solar.
The left-hand set of panels shows the region of the 
1425 index (a blend of Si, C, and Fe photospheric lines---see text), 
while the right-hand set is for the Fe{\footnotesize ~III} blend 
near 1978\,\AA. Each pair of panels is labelled with
the metallicity of the synthetic spectrum shown.
The $y$-axis is residual intensity (after normalization
by the continuum level).
\label{fig:lbg_models_cB58}}
\end{figure}
%%%%%%%%%%%%%%%%%%%%%%%%%%%%%%%%%%%%%%%%%%%%%%%%%%%%%%%%%%%%%%%%%%%%%%

\subsection{MS1512--cB58}\label{subsec:cb58_comp}

Thanks to gravitational lensing by the foreground cluster MS1512+36 at
$z=0.373$, MS1512--cB58 (or cB58 for short) is the brightest Lyman
break galaxy known and thus one of the most extensively studied.
Following its spectroscopic confirmation as a distant galaxy by
\cite{yee96} (the systemic redshift is $z_{\rm sys} = 2.7276$),
\cite{pettini00} analyzed spectra obtained with the Low Resolution
Imaging Spectrometer (LRIS) on the Keck telescope with a view to
probing the galaxy's physical and chemical properties. A more detailed
study of the kinematics and chemical abundances of the interstellar
gas in cB58 was made possible by high resolution and high S/N ratio
observations performed with ESI on Keck \citep{pettini02}. Here we
compare both the LRIS and ESI data with the synthetic spectra
generated by {\em Starburst99 + WM-basic\/}.

Since the observations of cB58 are of unusually high resolution (see
Table~\ref{tab:lbg_spectra}), we first degraded them to the FWHM =
2.5\,\AA\ of the synthetic spectra by convolution with Gaussian
profiles of the appropriate widths. The spectra were then normalised
by division by a spline fit to the continuum points listed in
Table~\ref{tab:continuum_windows} [omitting two regions near
1275\,\AA\ and 1750\,\AA\ where absorption lines from intervening
absorption systems, unrelated to cB58, were identified by
\cite{pettini00}].  We restrict our comparison between observed and
synthesized spectra to the wavelength regions of the 1425 and 1978
indices, and neglect the 1370 index which is compromised by lines from
two intervening absorption systems, \ion{Mg}{2}~$\lambda\lambda
2796,2804$ at $z_{\rm abs} = 0.8287$ and \ion{Si}{4}~$\lambda\lambda
1394,1403$ at $z_{\rm abs} = 2.6606$ \citep{pettini00, pettini02}.

Figure~\ref{fig:lbg_models_cB58} compares portions of the observed
spectrum with the corresponding theoretical spectra computed for five
metallicities in the 1400--1450\,\AA\ (left) and 1898--2025\,\AA\ 
(right) wavelength intervals.  We show the LRIS spectrum for the
former and the ESI one for the latter (although the ESI spectrum is in
general the more useful of the two for such a comparison, it does
suffer from poor flux calibration in the 1425\,\AA\ region where two
echellette orders overlap).  The first conclusion to be drawn is that
the {\em Starburst99 + WM-basic\/} combination does a remarkably good
job at synthesizing {\it ab initio} the general character of these
spectral regions, attesting to the high degree of sophistication
achieved by these codes.  Among the five metallicities considered, $Z
= 1 \Zsol$ and $Z = 0.4 \Zsol$ are those which most closely match the
observations in both the 1425 and 1978 regions (giving comparable rms
residuals between observed and model spectra).

A metallicity between \zsol\ and $0.4 \Zsol$ is in good agreement with
other measures.  Specifically, the rest-frame optical emission lines
analyzed by \cite{teplitz00} with the $R_{23}$ method of
\cite{pagel79} imply an oxygen abundance in the ionized gas of
$12+\log{\rm (O/H)}=8.39$, or $\sim 0.5$ of our reference `solar'
value $12+\log{\rm (O/H)}=8.72$ in Table~\ref{tab:orion_abundances}
for the Orion nebula. Although less precise, arguments based on the
strengths of the P-Cygni profiles of the \ion{C}{4}, \ion{Si}{4}, and
\ion{N}{5} wind lines also indicate a slightly sub-solar metallicity
\citep{pettini00, leitherer01}.

By far the most detailed assessment of element abundances in cB58
comes from the study of interstellar absorption lines from cool gas by
\cite{pettini02}. These authors found that among the elements covered
in the ESI spectrum of cB58, those which are promptly released into
the interstellar medium (Mg, Si, and S) all have abundances of $\sim
0.4$ solar with a scatter of about 0.1\,dex, comparable to the
measurement error. On the other hand, the Fe-peak elements they
observed have a mean metallicity $Z = \left ( 0.11^{+0.06}_{-0.04}
\right ) \Zsol$, that is they are less abundant by a further factor of
$\sim 4$, which \citeauthor{pettini02} attribute to a combination of
dust depletion and time delay in their release from Type~Ia
supernovae.  Evidently, the iron abundance in the OB stars, as deduced
here from the 1978 index, does not agree with the gas-phase abundance
of interstellar Fe determined by \cite{pettini02}, but is close to
those of interstellar Mg, Si, O, and S.  At face value, (Fe/H)$_{\rm
  OB~stars} \gtrsim 4 \times$~(Fe/H)$_{\rm gas}$.  This could be an
indication that most of the underabundance of interstellar Fe in cB58
is due to dust depletion---an interpretation not favoured by
\citeauthor{pettini02}, or that there is an offset in the calibration
of the 1978 index with metallicity, or a combination of both. An
additional contributing factor may be the unfortunate coincidence
that, at the redshift of cB58, the 1935--2020\,\AA\ wavelength region
is contaminated by telluric absorption; although this was divided out,
we could only do so approximately and the process may have left some
residual telluric absorption adding to the photospheric \ion{Fe}{3}
lines.

In summary, we conclude that the analysis of photospheric UV lines
from OB stars in cB58, using the methods developed here, yields an
estimate of metallicity which agrees with that of the interstellar
medium to within a factor of $\sim 2$ (allowing for a moderate degree
of dust depletion of the Fe-peak elements).

%%%%%%%%%%%%%%%%%%%%%%%%%%%%%%%%%%%%%%%%%%%%%%%%%%%%%%%%%%%%%%%%%%%%%%
\tabletypesize{\scriptsize}
\begin{deluxetable}{cccccc}
\tablewidth{238.53815pt}
\tablecaption{\textsc{Model Spectra for Continuous Star Formation with
  Salpeter IMF} \label{tab:model_spectra}}
\tablehead{
  \colhead{Wavelength} & \colhead{$Z = 0.05 Z_{\odot}$} & \colhead{$Z
    = 0.2 Z_{\odot}$} & \colhead{$Z = 0.4 Z_{\odot}$} & \colhead{$Z =
    1 Z_{\odot}$} & \colhead{$Z = 2 Z_{\odot}$}
}
\startdata
   1150.00   &  1.025  &   0.987   &  0.951   &  0.890   &  0.825\\
   1150.18   &  1.026  &   0.988   &  0.951   &  0.889   &  0.823\\
   1150.36   &  1.027  &   0.989   &  0.951   &  0.887   &  0.821\\
   1150.54   &  1.028  &   0.989   &  0.950   &  0.886   &  0.819\\
   1150.72   &  1.028  &   0.989   &  0.949   &  0.884   &  0.817\\
   1150.91   &  1.028  &   0.988   &  0.949   &  0.884   &  0.817\\
   1151.09   &  1.028  &   0.988   &  0.949   &  0.884   &  0.818\\
   1151.27   &  1.027  &   0.988   &  0.949   &  0.885   &  0.819\\
   1151.45   &  1.027  &   0.988   &  0.949   &  0.885   &  0.820\\
   1151.63   &  1.026  &   0.987   &  0.949   &  0.886   &  0.821\\
\enddata
\tablecomments{The units of wavelength are \AA. The other five columns
  list values of residual intensity at five different metallicities,
  as indicated in the column headings. Table~\ref{tab:model_spectra}
  is published in its entirety in the electronic version of the {\it
    Astrophysical Journal}. A portion is shown here for guidance
  regarding its form and content.}
\end{deluxetable}
%%%%%%%%%%%%%%%%%%%%%%%%%%%%%%%%%%%%%%%%%%%%%%%%%%%%%%%%%%%%%%%%%%%%%%

%%%%%%%%%%%%%%%%%%%%%%%%%%%%%%%%%%%%%%%%%%%%%%%%%%%%%%%%%%%%%%%%%%%%%%
\begin{figure}
\includegraphics[width=\columnwidth]{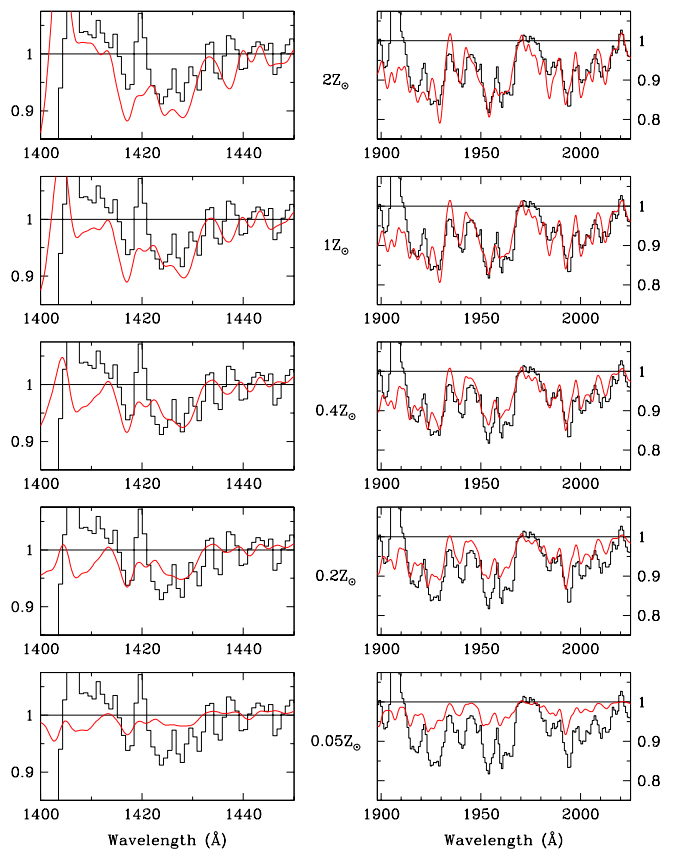}
\caption{Comparison of the LRIS-B spectrum 
  of the $z = 1.411$ galaxy Q1307--BM1163 from the survey by
  \cite{steidel04} ({\em black}) with {\em Starburst99 + WM-basic\/}
  synthetic spectra for five metallicities ({\em red}).  Other details
  are as in Figure~\ref{fig:lbg_models_cB58}.
\label{fig:lbg_models_BM1163}}
\end{figure}
%%%%%%%%%%%%%%%%%%%%%%%%%%%%%%%%%%%%%%%%%%%%%%%%%%%%%%%%%%%%%%%%%%%%%%

%%%%%%%%%%%%%%%%%%%%%%%%%%%%%%%%%%%%%%%%%%%%%%%%%%%%%%%%%%%%%%%%%%%%%%
\tabletypesize{\scriptsize}
\begin{deluxetable*}{lcccccc}
\tablewidth{472.51746pt}
\tablecaption{\textsc{Summary of Abundance Measurements in
    MS1512--cB58 and Q1307--BM1163}\label{tab:abun_summary}}
\tablehead{
  \colhead{Galaxy}
& \colhead{Redshift}
& \colhead{IS Abs. Lines\tablenotemark{a}}
& \colhead{$R_{23}$\tablenotemark{b}}
& \colhead{$N2$\tablenotemark{c}}
& \colhead{1425 index\tablenotemark{d}}
& \colhead{1978 index\tablenotemark{e}}
}
\startdata
MS1512--cB58  & 2.7276 & $Z = (0.4 \pm 0.1)\,\Zsol$ &
(O/H)\,=\,0.5\,(O/H)$_{\odot}$ \tablenotemark{f} & \nodata & $Z = 0.7\,\Zsol$ & (Fe/H)\,=\,0.7\,(Fe/H)$_{\odot}$ \\
              &        & (Fe/H)\,=\,$\left ( 0.11^{+0.06}_{-0.04} \right ) \times$\,(Fe/H)$_{\odot}$ \tablenotemark{g} &\\
\\
Q1307--BM1163 & 1.411\phn  & \nodata & \nodata & (O/H)\,$= \left (
  0.65^{+0.5}_{-0.3} \right ) \times  $\,(O/H)$_{\odot}$ & \nodata & (Fe/H)\,=\,1.3\,(Fe/H)$_{\odot}$ \\
\enddata             
\tablenotetext{a}{Abundances from interstellar absorption lines
  \citep{pettini02}.}
\tablenotetext{b}{Oxygen abundance in the ionized gas
  \citep{teplitz00}, deduced from the $R_{23}$ index ($R_{23}
  \equiv$\,\{[O{\scriptsize ~II}] + [O{\scriptsize ~III}]\}/H$\beta$) of
  \cite{pagel79}.}
\tablenotetext{c}{Oxygen abundance in the ionized gas
  \citep{steidel04}, deduced from the $N2$ index ($ N2
  \equiv$\,[N{\scriptsize ~II}]~$\lambda 6583$/H$\alpha$) of \cite{pp04}.}
\tablenotetext{d}{Metallicity of OB stars, deduced from the equivalent
  width of the photospheric 1425\,\AA\ blend
  [eq.~(\ref{eq:Z_EW_1425})].}
\tablenotetext{e}{Iron abundance of OB stars, deduced from the
  equivalent width of the photospheric 1978\,\AA\ blend
  [eq.~(\ref{eq:Z_EW_1978})].}
\tablenotetext{f}{Note that the measurement of \cite{teplitz00} is
  scaled to the value of (O/H)$_{\odot}$ adopted in the present work.
  \cite{teplitz00} do not provide an error estimate to the oxygen
  abundance. The typical accuracy of the $R_{23}$ method is a factor
  of $\sim 2$ \citep[e.g.][]{kennicutt03}.}
\tablenotetext{g}{This value is a lower limit, because an unknown
  fraction of the Fe-peak elements may be depleted onto interstellar
  grains.}
\end{deluxetable*}    
%%%%%%%%%%%%%%%%%%%%%%%%%%%%%%%%%%%%%%%%%%%%%%%%%%%%%%%%%%%%%%%%%%%%%%

\subsection{Q1307--BM1163}\label{subsec:BM1163_comp}

The combination of high luminosity and relatively low redshift makes
the ${\cal R} = 21.67$, $z = 1.411$ galaxy Q1307--BM1163 one of the
brightest in the BM/BX sample of galaxies in the `redshift desert' ($z
\simeq 1.4 - 2.5$) recently assembled by \cite{steidel04}.  These
authors reported an oxygen abundance $12 + \log {\rm (O/H)} = 8.53 \pm
0.25$ for the \ion{H}{2} gas from the observed [\ion{N}{2}]~$\lambda
6583$/H$\alpha$ emission line ratio \citep[the $N2$ index of][]{pp04};
this value corresponds to a metallicity $Z = \left (
  0.65^{+0.5}_{-0.3} \right ) \Zsol$ when compared with our reference
$12 + \log {\rm (O/H)} = 8.72$.  The profile of the \ion{C}{4} wind
line is also consistent with near-solar metallicity \citep[see Fig.~6
of][]{steidel04}.

Figure~\ref{fig:lbg_models_BM1163} compares the LRIS-B spectrum of
\cite{steidel04} with our five synthetic spectra, again in the
1400--1450\,\AA\ (left) and 1898--2025\,\AA\ (right) wavelength
intervals.  The 1360--1380\,\AA\ region is less useful in this case
because, at the redshift of Q1307--BM1163, it is close to the blue
edge of the LRIS-B spectrum, where the S/N is low.  As before, the
observations were suitably smoothed to match the resolution of the
model spectra (in this case only a moderate amount of smoothing was
required, from FWHM\,$= 2$\,\AA\ to FWHM\,$= 2.5$\,\AA) and normalised
by reference to the continuum points defined in
Table~\ref{tab:continuum_windows}.  Inspection of
Figure~\ref{fig:lbg_models_BM1163} indicates that the synthetic
spectrum with $Z = 0.4 \Zsol$ is the closest match (lowest rms
difference) to Q1307--BM1163 in the 1415--1435\,\AA\ region, while the
solar metallicity model reproduces best the \ion{Fe}{3} blend near
1978\,\AA. The latter may be more reliable, given the higher S/N of
the data at these wavelengths.  However, in either case the
conclusion, as for cB58, is that the photospheric UV stellar indices
give abundances that agree to within a factor of $\sim 2$ with those
deduced from other indicators based on stellar wind lines and nebular
emission lines from \ion{H}{2} regions.

\subsection{The Use of Line Indices}

\begin{figure}[b]
\includegraphics[angle=270,width=\columnwidth]{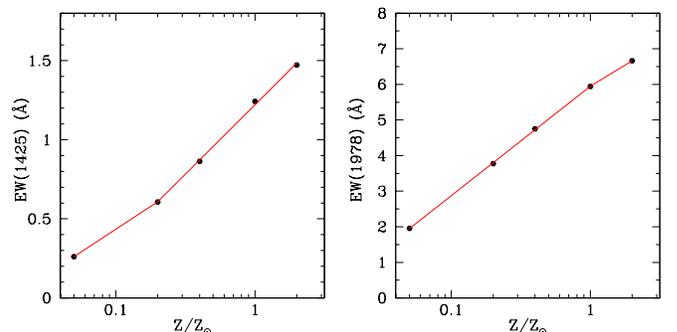}
\caption{EW$-Z$ relation for the 1425 and 1978 indices. The
dots are the values of equivalent width 
measured from the synthetic spectra 
at age 100\,Myr, while the lines show the functional fits
given by eqs.~(\ref{eq:Z_EW_1425}) and (\ref{eq:Z_EW_1978}).
\label{fig:plot_ew_Z_relation}}
\end{figure}

When we assessed the outputs of our code against the spectra of cB58
and Q1307--BM1163 in the preceding sections, we did so by identifying
the synthetic spectrum, among five options in metallicity, that gives
the smallest rms deviations from the observations.  For a more
straightforward determination of metallicity, we give here functional
forms to the relations between the equivalent widths of the 1425 and
1978 line blends and $\log (Z/\Zsol)$.  These were deduced by fitting
the values measured from our synthetic spectra at the five
metallicities for which they were generated, as shown in
Figure~\ref{fig:plot_ew_Z_relation}. The values of EW fitted are those
from Figures~\ref{fig:plot_ew_indices_1425} and
\ref{fig:plot_ew_indices_1978}, at time 100\,Myr. We find:
\begin{equation}\label{eq:Z_EW_1425}
\log (Z/\Zsol)=A \times \mbox{EW}(1425)+B,\quad \mbox{where}
\end{equation}
\begin{eqnarray}
A =1.74 \quad \mbox{and} \quad B=-1.75& & \quad \mbox{EW}(1425)<0.6\mbox{\AA} \nonumber\\
A =1.14 \quad \mbox{and} \quad B=-1.39& & \quad \mbox{EW}(1425) \geqslant 0.6\mbox{\AA} \nonumber
\end{eqnarray}

and

\begin{equation}\label{eq:Z_EW_1978}
\log (Z/\Zsol)=C \times \mbox{EW}(1978)+D,\quad \mbox{where}
\end{equation}
\begin{eqnarray}
C =0.33 \quad \mbox{and} \quad D=-1.94& & \quad \mbox{EW}(1978) \leqslant 5.9\mbox{\AA} \nonumber\\
C =0.42 \quad \mbox{and} \quad D=-2.50& & \quad \mbox{EW}(1978) > 5.9\mbox{\AA}.\nonumber
\end{eqnarray}

As can be seen from Figure~\ref{fig:plot_ew_Z_relation}, both indices
increase linearly, or nearly so, with $\log Z$ over the factor of 40
in $Z$ considered in this work. It is important to realize that the
relations given in (\ref{eq:Z_EW_1425}) and (\ref{eq:Z_EW_1978}) above
{\em are valid only for spectra of 2.5\,\AA\ resolution normalised
  through the continuum points listed in
  Table~\ref{tab:continuum_windows}\/}.  Given the broad, shallow
nature of these photospheric blends, their equivalent widths are very
sensitive to the continuum normalization which, as explained in
\S\ref{subsec:continuum_fit}, in turn depends on the spectral
resolution of the data.

%%%%%%%%%%%%%%%%%%%%%%%%%%%%%%%%%%%%%%%%%%%%%%%%%%%%%%%%%%%%%%%%%%%%%%
%%%%%%%%%%%%%%%%%%%%%%%%%%%%%%%%%%%%%%%%%%%%%%%%%%%%%%%%%%%%%%%%%%%%%%

Applying eqs.~(\ref{eq:Z_EW_1425}) and (\ref{eq:Z_EW_1978}) to cB58,
we find $\log Z = -0.16$ ($Z = 0.69 \Zsol$) from the measured
EW(1425)=1.08\,\AA, and $\log Z = -0.17$ ($Z = 0.68 \Zsol$) from
EW(1978)=5.37\AA.  Although in good agreement with each other, these
values are somewhat higher than $Z = (0.4 \pm 0.1) \Zsol$ determined
for the interstellar gas (\S\ref{subsec:cb58_comp}).  In Q1307--BM1163
we can only measure the 1978 index, because the 1425 blend is affected
by a noise spike---see Figure~\ref{fig:lbg_models_BM1163}.  (This is
one of the limitations in reducing a whole spectral feature to a
single number. While admittedly more objective, this approach is
however more vulnerable to noise than the spectral match discussed in
\S\ref{subsec:cb58_comp} and \S\ref{subsec:BM1163_comp}. With the
latter it is still possible to recover useful information from `clean'
portions of the spectrum).  We deduce $\log Z = +0.10$ ($Z = 1.3
\Zsol$) from the measured EW(1978)=6.2\,\AA\ in Q1307--BM1163, which
again is somewhat higher than $Z = \left ( 0.65^{+0.5}_{-0.3} \right )
\Zsol$ deduced from the [\ion{N}{2}]/H$\alpha$ emission line ratio
from \ion{H}{2} gas (\S\ref{subsec:BM1163_comp}).

All the above abundance measurements in cB58 and Q1307--BM1163 are
summarised in Table~\ref{tab:abun_summary}.  Considering that each of
the established methods (interstellar absorption lines, $R_{23}$, and
$N2$) give values of metallicity which are only accurate to within a
factor of $\sim 2$, the photospheric line indices developed here
appear to be a promising alternative.  On the basis of this very
limited comparison, it would seem that the hot star models tend to
slightly underestimate the strengths of the photospheric blends at
1425\,\AA\ and 1978\,\AA\ at a given metallicity. It would be
interesting, in future, to verify if this is indeed the case with a
larger set of galaxies where both stellar and \ion{H}{2} region
abundances have been measured.

%%%%%%%%%%%%%%%%%%%%%%%%%%%%%%%%%%%%%%%%%%%%%%%%%%%%%%%%%%%%%%%%%%%%%%
\begin{deluxetable}{cccc}
\tablewidth{182.63141pt}
\tablecaption{\textsc{IMF Parameters}\label{tab:imf_par}}
\tablehead{
  \colhead{Model}
& \colhead{Star Formation Mode}
& \colhead{IMF slope\tablenotemark{a}}
& \colhead{$M_{\rm up}~(M_{\odot})$\tablenotemark{b}}
}
\startdata
A  &  Continuous     &  2.35  &      100 \\
B  &  Continuous     &  3.30  &      100 \\
C  &  Continuous     &  2.35  &  \phn 30 \\
\enddata             
\tablenotetext{a}{The parameter $\alpha$ in eq.~(\ref{eq:imf}).}
\tablenotetext{b}{Upper mass limit}
\end{deluxetable}    
%%%%%%%%%%%%%%%%%%%%%%%%%%%%%%%%%%%%%%%%%%%%%%%%%%%%%%%%%%%%%%%%%%%%%%

%%%%%%%%%%%%%%%%%%%%%%%%%%%%%%%%%%%%%%%%%%%%%%%%%%%%%%%%%%%%%%%%%%%%%%
\begin{figure}[b]
\includegraphics[angle=270,width=\columnwidth]{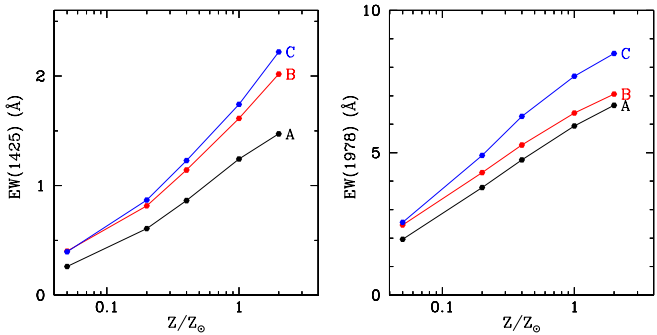}
\caption{Dependence of the 1425 and 1978 indices on the 
IMF parameters. 
Model A is for our standard case of continuous
star formation with a Salpeter IMF at age 100\,Myr.
Models B and C have comparatively fewer massive
stars than model A (see Table~\ref{tab:imf_par}).
The lines here simply join the dots, to guide the eye.
\label{fig:plot_ew_imf_variation}}
\end{figure}
%%%%%%%%%%%%%%%%%%%%%%%%%%%%%%%%%%%%%%%%%%%%%%%%%%%%%%%%%%%%%%%%%%%%%%

\subsubsection{Sensitivity of the Line Indices to the IMF parameters}\label{sec:imf_comparison}

Finally, we briefly consider the sensitivity of the 1425 and 1978
indices to the IMF parameters.  To this end, we generated a new set of
synthetic spectra as described in \S\ref{sec:synthetic_spectra}, but
with different IMF prescriptions, and remeasured values of EW(1425)
and EW(1978) at the five metallicities considered in this work.
Table~\ref{tab:imf_par} gives details of the IMF parameters
considered.  Model `A' is our standard case of continuous star
formation with a Salpeter IMF at age 100\,Myr.  Model `B' has a
steeper power-law slope ($\alpha = 3.30$ in eq.~\ref{eq:imf}), and is
thus close to the \cite{miller-scalo79} IMF at the high mass end.
Model `C' has the standard Salpeter slope but the IMF is truncated at
$M_{\rm up} = 30\,M_{\odot}$. Thus, both models B and C have
comparatively fewer high mass stars than model A.

\begin{figure}
\includegraphics[angle=270,width=\columnwidth]{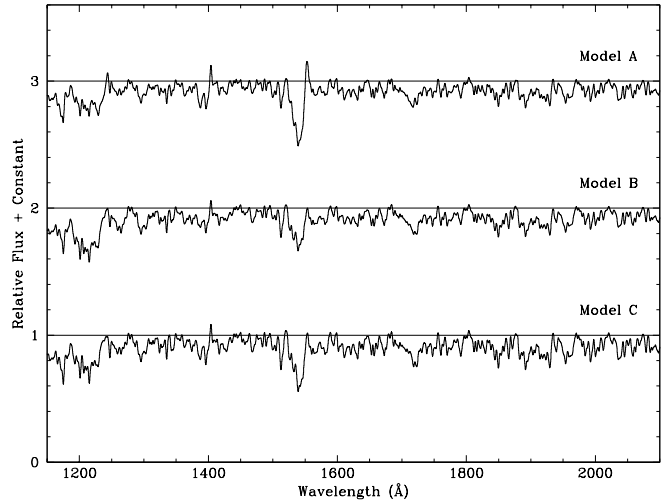}
\caption{Normalised, synthetic UV spectra at solar metallicity 
for the three sets of IMF parameters listed in Table~\ref{tab:imf_par}. 
Model A is for our standard case of continuous
star formation with a Salpeter IMF at age 100\,Myr.
Models B and C have comparatively fewer massive
stars than model A. 
\label{fig:plot_spectra_imf_variation}}
\end{figure}

As can be seen from Figure~\ref{fig:plot_ew_imf_variation}, decreasing
the proportion of massive stars consistently increases the equivalent
widths of the two blends (at a given metallicity), reflecting the fact
that these spectral features are strongest in late O and early B-type
stars. However, there are other spectral changes which accompany these
IMF alterations. The most obvious one (see
Figure~\ref{fig:plot_spectra_imf_variation}) is in the character of
the \ion{C}{4}~$\lambda \lambda 1548, 1550$ line which in models B and
C no longer exhibits a discernible P-Cygni component, as already noted
by \cite{pettini00} and \cite{steidel04}.  The empirical observation
that Lyman break galaxies normally show such P-Cygni features---as
demonstrated by the composite spectrum of \cite{shapley03}---argues
against significant departures from a Salpeter IMF (at least for
masses $M > 5\,M_{\odot}$) even at these early epochs.

%%%%%%%%%%%%%%%%%%%%%%%%%%%%%%%%%%%%%%%%%%%%%%%%%%%%%%%%%%%%%%%%%%%%%%
%%%%%%%%%%%%%%%%%%%%%%%%%%%%%%%%%%%%%%%%%%%%%%%%%%%%%%%%%%%%%%%%%%%%%%

\section{DISCUSSION AND CONCLUSIONS}\label{sec:conclusions}

Rest-frame UV spectra of galaxies at redshifts $z > 1$ are now being
routinely obtained with large ground-based telescopes.  Current
samples already include thousands of galaxies with spectroscopic
redshifts and their numbers will continue to grow in the years ahead.
Encoded in all these spectra is information on the chemical
composition of the galaxies which is one of the key physical
properties for understanding their previous history and future
evolution. Potentially, knowledge of the relative abundances of the
chemical elements in these galaxies can tell us about the accumulation
of the products of previous stellar generations, the consumption of
gas into stars, the pace of star formation activity, the initial mass
function, and the importance of inflows and outflows in shaping the
chemical evolution of galaxies in the distant past.

In this paper we have taken the first steps towards decoding some of
this information. In order to do so, we have found it necessary to
synthesize {\em ab initio\/} the spectra of stars at the upper end of
the H-R diagram and use them, via a spectral synthesis code, to
predict the emergent UV spectrum of a composite stellar population as
a function of metallicity. This approach was dictated by the practical
difficulties with current facilities to assemble such a library of
stellar spectra directly from observations and was made possible by
the high degree of sophistication achieved by codes such as {\em
  WM-basic\/} and {\em Starburst99\/}.  A further advantage of a fully
theoretical method such as ours is that one has complete control over
the wavelength coverage and resolution of the synthesized spectrum, as
well as over the metallicity and relative abundances of individual
elements.

The results of this initial study are encouraging, in that theoretical
and empirical spectra resemble each other closely and in
detail---while the models may not be perfect yet, they provide a
remarkably good representation of reality.

We have focussed our attention on three blends of photospheric stellar
lines, near 1370, 1425, and 1978\,\AA, and found that they are indeed
mostly sensitive to metallicity.  In the idealized case of a
continuous star formation activity at a constant rate, the equivalent
widths of these three features vary little with time after
approximately 50\,Myr, and increase by factors of $\sim 4 - 6$ as the
metallicity of the stars increases from 1/20 of solar to twice solar.
Of the three blends, the one at 1978\,\AA\ may be the most useful
because: (a) it is the strongest, with an equivalent width EW(1978)\,$
= 5.9$\,\AA\ at solar metallicity; (b) the wavelength interval
1935--2020\,\AA\ over which it is defined is free of strong
interstellar lines; (c) its rest-frame wavelength makes it accessible
to ground-based spectroscopy over the redshift range $z = 1 - 3$ where
recent surveys for high redshift galaxies have been prolific; and (d)
it is due primarily to \ion{Fe}{3} lines and thus measures the
abundance of a single element rather than a combination of several.

We have only been able to test the validity of our method in two
cases, by comparing the synthetic spectra we generated with those of
two bright high redshift galaxies, MS1512--cB58 at $z = 2.7276$ and
Q1307--BM1163 at $z = 1.411$. We find that in these two cases the
metallicities deduced from the photospheric lines agree, to within a
factor of $\sim 2$, with those derived from analyses of interstellar
absorption and emission lines. More extensive comparisons will allow
us to refine the calibrations of our line indices in terms of
metallicity.

The principal conclusion of this work is that photospheric UV lines
offer a much needed complementary avenue to determining the degree of
metal enrichment in high redshift galaxies. The main weaknesses of the
method we have developed are: (a) It is based on spectral features
that are broad and shallow. Thus, while they can be detected even in
spectra of modest resolution, their measurement does require a
moderately high signal-to-noise ratio, S/N\,$\gtrsim 15$ per pixel at
$R \simeq 1000$.  (b) The equivalent widths of the features in
question depend on the mix of spectral types present in a star-forming
region.  Thus, their strengths would vary with time in isolated bursts
of star formation and, to a lesser extent, with the initial mass
function.  However, to date, neither single bursts nor departures from
a universal IMF (at the upper mass end) have been found to be common
in high redshift galaxies.

The main strength of the technique is its complementarity to other
abundance measures: it is applicable precisely where other techniques
fall short. Abundance determinations from interstellar absorption
lines require a combination of spectral resolution and S/N that is not
achievable with current means for the vast majority of high redshift
galaxies.  Circumventing this problem by co-adding spectra of many
galaxies is fraught with dangers \citep[e.g.][]{savaglio04}, since the
absorption lines are normally on the flat part of the curve of growth,
where their equivalent widths are much more sensitive to the velocity
dispersion of the gas than to its column density (and therefore
metallicity).  Furthermore, most abundant elements are depleted onto
dust in the interstellar medium; correcting for unknown and variable
amounts of dust depletion is a significant complication which has not
yet been fully resolved even in QSO absorption line studies
\citep[e.g.][]{vladilo04}.

Emission lines from \ion{H}{2} regions, on the other hand, when
observed at high redshifts are limited to the strongest lines which
sample only a few elements (primarily oxygen).  The so-called `strong
line methods' give estimates of the oxygen abundance that are only
accurate to within a factor of $\sim 2$ [see, for example, the
discussion by \cite{pp04}], comparable to the accuracy of the
photospheric UV line method explored here, at least on the basis of
initial tests.  Furthermore, there are only a few redshifts intervals
at $z > 1$ where the nebular lines required for abundance measurements
all fall within near-IR windows accessible from the ground, and even
then contamination by strong sky emission lines is often a limiting
factor to the measurement of reliable line ratios.

Looking ahead, there are thus strong motivations for continuing the
work we have begun by: (a) incorporating future improvements in {\em
  WM-basic\/} into our hybrid code and, (b) most importantly,
performing more extensive cross-checks between abundances deduced from
nebular emission lines and with our photospheric UV absorption line
technique, in cases where both methods are applicable.  With a
reliable calibration of the latter in hand, we would then have the
means to measure stellar abundances in a wholesale manner over a wide
range of cosmic epochs.

\acknowledgements 
We would like to express our gratitude to Adi Pauldrach and his
collaborators for making the {\em WM-basic\/} code freely available
and for their kind assistance in the initial stages of this project.
This work has benefited from numerous discussions with Paul Crowther,
Chris Evans, Danny Lennon, Stephen Smartt, and Carrie Trundle; Alice
Shapley and the anonymous referee made a number of suggestions which
improved the presentation of the paper.  S. Rix is grateful to the
Isaac Newton Group of Telescopes on La Palma for their hospitality
during the writing of this paper, and to the UK Particle Physics and
Astronomy Research Council for the postgraduate studentship which
supported her research.

%%%%%%%%%%%%%%%%%%%%%%%%%%%%%%%%%%%%%%%%%%%%%%%%%%%%%%%%%%%%%%%%%%%%%%

% Add the bibliography.

\clearpage

%%%%%%%%%%%%%%%%%%%%%%%%%%%%%%%%%%%%%%%%%%%%%%%%%%%%%%%%%%%%%%%%%%%%%%

% Add the long table at the end of the document.
% Explicitly give the number of the table as Table 1.

\LongTables
\begin{deluxetable}{lcccccc}
\tablewidth{0pt}
\tablenum{1}
\label{tab:wmb_grid} % Positioned after \tablenum as recommended.
\tablecaption{\textsc{Grid of WM-Basic Models}}
\tablehead{
  \colhead{$M$} 
& \colhead{$\log$ ($L$/$\Lsol$)}
& \colhead{$T_{\rm eff}$}
& \colhead{$\log$ ($g$/cm~s$^{-2}$)}
& \colhead{$R$}
& \colhead{\mdot}
& \colhead{$v_{\infty}$}\\
  \colhead{($\Msol$)} 
& \colhead{}
& \colhead{(K)}
& \colhead{}
& \colhead{($\Rsol$)}
& \colhead{($\Msol$ yr$^{-1}$)}
& \colhead{(\kms)}
}
\startdata
      119.98 & 6.25 &  59870 & 4.333 &     12.39 &   0.18E-04 &     4076  \\*
      119.61 & 6.25 &  52061 & 4.091 &     16.34 &   0.17E-04 &     3545  \\*  
      113.24 & 6.25 &  49139 & 3.965 &     18.39 &   0.31E-04 &     3190  \\*     
      103.40 & 6.25 &  46967 & 3.850 &     20.06 &   0.33E-04 &     2829  \\*     
\phm{1}91.79 & 6.25 &  46330 & 3.775 &     20.59 &   0.36E-04 &     2498  \\*     
\phm{1}74.11 & 6.23 &  25085 & 2.636 &     68.66 &   0.64E-05 &     1187  \\*     
\phm{1}35.76 & 5.90 &  22339 & 2.449 &     59.17 &   0.85E-06 &  \phn837  \\*     
\phm{11}9.15 & 4.93 &  25100 & 3.025 &     15.43 &   0.26E-06 &     1131  \\     
\\                                                         
\phm{1}99.63 & 6.13 &  58567 & 4.334 &     11.27 &   0.11E-04 &     3997  \\*
\phm{1}99.75 & 6.13 &  50928 & 4.094 &     14.87 &   0.11E-04 &     3489  \\*     
\phm{1}96.50 & 6.14 &  48498 & 3.982 &     16.65 &   0.12E-04 &     3187  \\*    
\phm{1}91.23 & 6.15 &  46122 & 3.859 &     18.64 &   0.22E-04 &     2854  \\*     
\phm{1}83.46 & 6.16 &  43501 & 3.712 &     21.13 &   0.25E-04 &     2466  \\*     
\phm{1}70.95 & 6.16 &  35514 & 3.289 &     31.71 &   0.30E-04 &     1724  \\*     
\phm{1}51.76 & 6.08 &  31526 & 3.020 &     36.90 &   0.28E-04 &     1236  \\*     
\phm{1}13.17 & 5.35 &  33117 & 3.249 &     14.30 &   0.20E-05 &     1235  \\     
\\   
\phm{1}80.13 & 5.96 &  57000 & 4.363 &  \phn9.79 &   0.57E-05 &     4011  \\*
\phm{1}79.86 & 5.96 &  49566 & 4.123 &     12.87 &   0.56E-05 &     3497  \\*     
\phm{1}78.64 & 5.98 &  47765 & 4.032 &     14.19 &   0.63E-05 &     3262  \\*     
\phm{1}76.50 & 6.00 &  45682 & 3.922 &     15.89 &   0.71E-05 &     2986  \\*     
\phm{1}72.72 & 6.01 &  42474 & 3.755 &     18.75 &   0.14E-04 &     2615  \\*    
\phm{1}65.85 & 6.03 &  38434 & 3.526 &     23.24 &   0.16E-04 &     2145  \\*     
\phm{1}57.50 & 6.05 &  36653 & 3.363 &     26.20 &   0.20E-04 &     1749  \\*     
\phm{1}37.09 & 6.01 &  24041 & 2.476 &     58.40 &   0.54E-05 &  \phn742  \\     
\\                                                      
\phm{1}60.07 & 5.73 &  54635 & 4.394 &  \phn8.17 &   0.23E-05 &     3968  \\*
\phm{1}59.94 & 5.73 &  47509 & 4.150 &     10.82 &   0.24E-05 &     3444  \\*     
\phm{1}59.43 & 5.75 &  46277 & 4.078 &     11.70 &   0.26E-05 &     3266  \\*     
\phm{1}58.68 & 5.78 &  44963 & 3.997 &     12.76 &   0.30E-05 &     3071  \\*     
\phm{1}57.54 & 5.80 &  43138 & 3.891 &     14.28 &   0.34E-05 &     2832  \\*     
\phm{1}55.65 & 5.83 &  40319 & 3.734 &     16.82 &   0.40E-05 &     2517  \\*     
\phm{1}52.28 & 5.85 &  34971 & 3.436 &     22.98 &   0.80E-05 &     2028  \\*     
\phm{1}46.34 & 5.88 &  25314 & 2.798 &     45.11 &   0.24E-05 &     1308  \\     
\\                                                      
\phm{1}49.97 & 5.57 &  52465 & 4.403 &  \phn7.37 &   0.13E-05 &     3906  \\*
\phm{1}49.96 & 5.57 &  45622 & 4.162 &  \phn9.73 &   0.12E-05 &     3401  \\*     
\phm{1}49.23 & 5.62 &  43576 & 4.029 &     11.27 &   0.16E-05 &     3089  \\*     
\phm{1}47.81 & 5.67 &  40507 & 3.837 &     13.85 &   0.20E-05 &     2684  \\*     
\phm{1}46.55 & 5.69 &  37807 & 3.679 &     16.39 &   0.24E-05 &     2396  \\*     
\phm{1}44.20 & 5.72 &  33037 & 3.395 &     22.14 &   0.48E-05 &     1964  \\*    
\phm{1}39.75 & 5.75 &  24880 & 2.826 &     40.41 &   0.15E-05 &     1331  \\*     
\phm{1}16.89 & 5.59 &  29759 & 2.925 &     23.51 &   0.19E-05 &  \phn890  \\     
\\                                                      
\phm{1}39.94 & 5.38 &  50294 & 4.423 &  \phn6.45 &   0.60E-06 &     3820  \\*
\phm{1}39.97 & 5.38 &  43734 & 4.184 &  \phn8.49 &   0.59E-06 &     3333  \\*     
\phm{1}39.50 & 5.42 &  42084 & 4.067 &  \phn9.66 &   0.73E-06 &     3072  \\*     
\phm{1}38.78 & 5.47 &  40205 & 3.928 &     11.23 &   0.94E-06 &     2780  \\*     
\phm{1}37.45 & 5.53 &  36500 & 3.691 &     14.49 &   0.12E-05 &     2354  \\*     
\phm{1}36.28 & 5.56 &  32858 & 3.465 &     18.51 &   0.14E-05 &     2017  \\*     
\phm{1}34.37 & 5.59 &  26494 & 3.035 &     29.55 &   0.10E-05 &     1521  \\*     
\phm{1}15.70 & 5.53 &  32953 & 3.132 &     17.87 &   0.46E-05 &     1037  \\
\\
\phm{1}29.95 & 5.11 &  45971 & 4.412 &  \phn5.65 &   0.45E-07 &     3624  \\*
\phm{1}29.99 & 5.11 &  39975 & 4.172 &  \phn7.46 &   0.40E-07 &     3159  \\*     
\phm{1}29.33 & 5.19 &  37688 & 3.975 &  \phn9.25 &   0.69E-07 &     2763  \\*     
\phm{1}28.47 & 5.27 &  34549 & 3.733 &     12.04 &   0.45E-06 &     2343  \\*     
\phm{1}28.02 & 5.30 &  32589 & 3.595 &     14.01 &   0.52E-06 &     2136  \\*     
\phm{1}27.43 & 5.33 &  29727 & 3.394 &     17.46 &   0.20E-06 &     1866  \\*     
\phm{1}26.55 & 5.37 &  24838 & 3.032 &     26.07 &   0.58E-06 &     1480  \\*     
\phm{1}15.62 & 5.45 &  15267 & 1.879 &     75.41 &   0.30E-06 &  \phn303  \\
\\ 
\phm{1}24.97 & 4.91 &  43811 & 4.449 &  \phn4.95 &   0.11E-07 &     3592  \\*
\phm{1}24.99 & 4.91 &  38097 & 4.211 &  \phn6.50 &   0.11E-07 &     3135  \\*     
\phm{1}24.61 & 4.97 &  36524 & 4.062 &  \phn7.66 &   0.17E-07 &     2842  \\*     
\phm{1}24.35 & 5.02 &  35524 & 3.969 &  \phn8.49 &   0.23E-07 &     2670  \\*    
\phm{1}24.02 & 5.06 &  34052 & 3.842 &  \phn9.75 &   0.33E-07 &     2455  \\*     
\phm{1}23.81 & 5.09 &  32982 & 3.758 &     10.70 &   0.39E-07 &     2323  \\*     
\phm{1}23.25 & 5.14 &  29531 & 3.499 &     14.25 &   0.59E-07 &     1952  \\*     
\phm{1}22.85 & 5.18 &  26636 & 3.279 &     18.19 &   0.75E-07 &     1699  \\*     
\phm{1}22.41 & 5.21 &  23768 & 3.040 &     23.73 &   0.38E-06 &     1458  \\
\\                                                       
\phm{1}19.96 & 4.66 &  40422 & 4.462 &  \phn4.36 &   0.21E-08 &     3468  \\*       
\phm{1}19.99 & 4.66 &  35150 & 4.223 &  \phn5.74 &   0.21E-08 &     3024  \\*   
\phm{1}19.85 & 4.70 &  34298 & 4.137 &  \phn6.32 &   0.28E-08 &     2864  \\*     
\phm{1}19.61 & 4.76 &  33160 & 4.008 &  \phn7.28 &   0.44E-08 &     2636  \\*     
\phm{1}19.28 & 4.84 &  31150 & 3.812 &  \phn9.04 &   0.78E-08 &     2324  \\*     
\phm{1}19.06 & 4.89 &  29320 & 3.655 &     10.78 &   0.11E-07 &     2086  \\*     
\phm{1}18.92 & 4.92 &  27925 & 3.540 &     12.26 &   0.13E-07 &     1941  \\*     
\phm{1}18.54 & 4.98 &  24035 & 3.205 &     17.85 &   0.20E-06 &     1574  \\     
\\                                                              
\phm{1}14.98 & 4.31 &  35864 & 4.479 &  \phn3.70 &   0.22E-09 &     3302  \\*
\phm{1}15.00 & 4.31 &  31186 & 4.236 &  \phn4.90 &   0.22E-09 &     2871  \\*     
\phm{1}14.91 & 4.37 &  30294 & 4.120 &  \phn5.58 &   0.34E-09 &     2674  \\*     
\phm{1}14.88 & 4.40 &  30016 & 4.079 &  \phn5.85 &   0.41E-09 &     2608  \\*     
\phm{1}14.84 & 4.43 &  29660 & 4.030 &  \phn6.18 &   0.49E-09 &     2503  \\*     
\phm{1}14.80 & 4.45 &  29252 & 3.976 &  \phn6.56 &   0.60E-09 &     2421  \\*     
\phm{1}14.75 & 4.48 &  28766 & 3.915 &  \phn7.03 &   0.74E-09 &     2332  \\*     
\phm{1}14.60 & 4.57 &  26393 & 3.671 &  \phn9.27 &   0.13E-08 &     2008  \\*     
\phm{1}14.50 & 4.63 &  24266 & 3.467 &     11.68 &   0.21E-08 &     1774  \\*
\phm{1}14.43 & 4.67 &  22909 & 3.325 &     13.72 &   0.27E-08 &     1599  \\
\\                                                       
\phm{1}10.00 & 3.77 &  25453 & 4.250 &  \phn3.94 &   0.31E-10 &     2611  \\*    
\phm{11}9.99 & 3.88 &  24487 & 4.074 &  \phn4.82 &   0.31E-10 &     2355  \\*     
\phm{11}9.99 & 3.89 &  24347 & 4.048 &  \phn4.97 &   0.31E-10 &     2319  \\     
\\
\phm{11}5.00 & 2.74 &  17169 & 4.292 &  \phn2.65 &   0.83E-10 &     1226  \\     
\enddata
\end{deluxetable}
{\footnotesize \noindent {\sc Note ---} The wind parameters \mdot\ and
  $v_{\infty}$ refer to the values used for models at solar
  metallicities. For the remaining metallicities, these parameters
  were scaled according to eqs.~(\ref{eq:v_term_Z}) and
  (\ref{eq:m_dot_Z})}.

%%%%%%%%%%%%%%%%%%%%%%%%%%%%%%%%%%%%%%%%%%%%%%%%%%%%%%%%%%%%%%%%%%%%%%

\end{document}